\documentclass[10pt]{iopart}
\usepackage[dvips]{graphicx}
\usepackage{amsthm}
\usepackage{amssymb}
\usepackage{bm}
\usepackage{amstext}
\usepackage{psfrag}
\usepackage{amsfonts}
\usepackage{array}

\newcommand{\p}{\partial}

\newcommand{\dd}{{\rm d}}

\newcommand{\bd}{\begin{definition}}                
\newcommand{\ed}{\end{definition}}                  
\newcommand{\bc}{\begin{corollary}}                 
\newcommand{\ec}{\end{corollary}}                   
\newcommand{\bl}{\begin{lemma}}                     
\newcommand{\el}{\end{lemma}}                       
\newcommand{\bp}{\begin{proposition}}            
\newcommand{\ep}{\end{proposition}}                
\newcommand{\bere}{\begin{remark}}                  
\newcommand{\ere}{\end{remark}}                     
\newtheorem{theorem}{Theorem}[section]
\newtheorem{corollary}[theorem]{Corollary}
\newtheorem{lemma}[theorem]{Lemma}
\newtheorem{proposition}[theorem]{Proposition}
\theoremstyle{definition}
\newtheorem{definition}[theorem]{Definition}
\theoremstyle{remark}
\newtheorem{remark}[theorem]{Remark}

\begin{document}

\title[Eisenhart's theorem and the causal simplicity of Eisenhart's spacetime]{Eisenhart's theorem and the causal simplicity \\ of Eisenhart's spacetime}

\author{E Minguzzi }

\address{Department of Applied Mathematics, Florence
 University, Via S. Marta 3, 50139 Florence, Italy}
\ead{ettore.minguzzi@unifi.it}
\begin{abstract}
We give a causal version of Eisenhart's geodesic characterization of
classical mechanics. We emphasize  the geometric, coordinate
independent properties needed to express Eisenhart's theorem in
light of modern studies on the Bargmann structures (lightlike
dimensional reduction, pp-waves). The construction of the space
metric, Coriolis 1-form and scalar potential through which the
theorem is formulated is shown in detail, and in particular it is
proved a one-to-one correspondence between Newtonian frames and
Abelian connections on suitable lightlike principal bundles. The
relation of Eisenhart's theorem in the lightlike case  with a Fermat
type principle is pointed out. The operation of lightlike lift is
introduced and the existence of minimizers for the classical action
is related to the causal simplicity of Eisenhart's spacetime.

\end{abstract}

%
%
%
%
%


\section{Introduction}
The mathematical study of lightlike dimensional reduction began
almost unnoticed in 1929 when Eisenhart published a work
\cite{eisenhart29} which drew a correspondence between the
trajectories of a dynamical system on a classical configuration
space $E$ (extended to include the time) and the geodesics of a
higher dimensional Lorentzian manifold $M$. While this result
attracted some interest, as it was going in the direction of
geometrizing Newtonian mechanics as for the older Jacobi metric
model \cite{arnold78},  the nature of the projection $\pi: M \to E$
and the lightlike character of the orbits $\pi^{-1}(e)$, $e \in E$,
was not considered as an interesting ingredient of the construction
\cite{lichnerowicz55}.

Eisenhart's results in connection with Lorentzian geometry were then
overlooked and only later rediscovered following different
approaches \cite{duval85,duval91}. Indeed, even the relation between
Eisenhart's result and lightlike dimensional reduction is not well
known. The aim of this work is to generalize Eisenhart's theorem is
such a way that it can become an integral part of present
investigations on the mathematics of lightlike dimensional
reduction.

It must be mentioned that the class of Eisenhart's spacetimes is
quite large and includes, for instance, the plane-fronted waves
considered in general relativity \cite{ehlers62}. Any result
obtained for the Eisenhart's spacetimes specializes immediately to
those. Although the terminology regarding plane waves is not always
uniform we may define, following \cite{ehlers62}, a {\em
plane-fronted wave} as (i) a 4-dimensional spacetime  which (ii)
satisfies Einstein equations  (most authors consider only the vacuum
case) and that (iii) admits a lightlike twist-free Killing field,
plus (iv) some auxiliary topological and differentiable conditions.
As we shall see, without the limit on the dimension of the spacetime
given by (i) and the condition of solving the Einstein equations
(ii), this definition would be that of an Eisenhart spacetime. If
the condition that (v) the Killing field is covariantly constant is
added then the class of {plane-fronted waves} restricts itself to
that of the {\em plane-fronted waves with parallel rays}, i.e. {\em
pp-waves}, whereas the Eisenhart's spacetime becomes a {\em Bargmann
structure}.

It seems appropriate to keep the name {\em Eisenhart's spacetime}
instead of {\em plane-fronted wave} for the structures to be studied
in this work not only for historical precedence or because of the
mentioned differences, but also for a fundamental difference in the
motivations. Eisenhart's spacetimes serve as a tool for
investigating the relationship between classical and relativistic
physics, and thus they are not meant as realistic spacetimes, rather
they provide a mean for mirroring  classical spacetimes into
relativistic ones.

As we shall see, the Eisenhart's spacetime is interesting for two
main reasons. On the one hand because it provides a bridge between
relativistic and non-relativistic physics and allows to export
methods and ideas from one subject to the other (see theorem
\ref{cas}). On the other hand because several physical interesting
spacetimes are indeed of Eisenhart type, thus for instance, the
results obtained in this work can be applied to the plane-fronted
waves of  general relativity.

Let $E= T\times S$, $T=\mathbb{R}$, be a classical $d+1$-dimensional
extended configuration space of coordinates $(t,q)$. Let $a_t$ be a
positive definite time dependent metric on $S$, and $b_t$ a time
dependent 1-form field on $S$. Let $\mathcal{U}(t,q)$ be a time
dependent scalar field on $S$. In a coordinate chart of $S$, we
shall write $a_t=a_{ab}\dd q^a\dd q^b$, $b_t= b_c\dd q^c $.
Eisenhart \cite{eisenhart29} was able to show that the trajectories
of a Lagrangian system with $d$ degrees of freedom
\begin{equation} \label{clas}
L(q, \dot{q},t)=\frac{1}{2}a_t(\dot{q},\dot{q})
+b_t(\dot{q})-\mathcal{U}(t,q) ,
\end{equation}
may be obtained as the projection of the geodesics of a
$d+2$-dimensional manifold $M=E\times Y$, $Y= \mathbb{R}$, of metric
\cite{eisenhart29,lichnerowicz55} \footnote{In our convention the
roles of $q^0$ and $q^{d+1}$ are inverted with respect to \cite[Book
II, Sect. 11]{lichnerowicz55} and there is also a different choice
of sign.}
\begin{eqnarray} \label{eis}
\dd s^2&=&a_t -\dd t \otimes (\dd y-b_t) -(\dd y-b_t) \otimes \dd
t-2(\mathcal{U}+k) \dd t ^2 ,
\end{eqnarray}
where $k\in \mathbb{R}$ is a constant, $y \in \mathbb{R}$ is the
variable of the additional $(d+1)$th dimension ($y=q^{d+1}$), and
$t$ is the zeroth dimension $t=q^{0}$. Different values of $k$ do
not really give different spacetimes, indeed $k$ can be sent to zero
through the transformations $y=y'-k t$. Thus the chosen value of $k$
selects a coordinate system rather than a metric, and the
possibility of choosing $k$  reflects the freedom in choosing the
zero level for the potential. Note that $\p/\p y$ is a lightlike
Killing vector field.

 Eisenhart did not gave to the Lorentzianity \footnote{The
Lorentzianity of the metric follows immediately by introducing the
base of 1-forms $\omega^{0}=\dd t $, $\omega^a=\dd q^a$,
$\omega^{d+1}=\dd y+(\mathcal{U}+k)\dd t -b$. } of his manifold a
particular meaning, and indeed, this fact played  no particular role
in his analysis. Given a solution of the dynamical system $q(t)$, he
set
\begin{equation} \label{dd}
y(t)=C- (\frac{1}{2}+k) t+\int_0^t L(q,\dot{q},t) \dd t ,
\end{equation}
where $C$ is an arbitrary constant. The trajectory $\{t, q(t),
y(t)\}$ could then be regarded as a parametrization of a {\em
spacelike} geodesic of the Eisenhart spacetime with respect to a
natural parameter \cite{lichnerowicz55}, $s=t$. Moreover, every
solution of the Lagrangian system could be regarded in this way.

A first question is whether Eisenhart's spacetime can be
characterized in some coordinate independent way and whether
Eisenhart's theorem  admits a {\em causal} formulation so that  the
representing geodesics become {\em causal} rather than {\em
spacelike}. An affirmative answer to the last question would give to
the causal geodesics the physical interpretation of particles moving
on Eisenhart's spacetime. In fact, we shall show that both questions
admit an affirmative answer and that the causal version of
Eisenhart's theorem is  convenient from the mathematical point of
view as it allows us to obtain new results on the existence of
minimizers for the classical action.

The modern coordinate-independent approach to lightlike dimensional
reduction is the so called {\em Bargmann structure} approach
\cite{duval85,duval91}, thus our aim will be to start with such
structures  or with the slight generalization considered in
\cite{julia95}, and then to recover the manifold $E$, the space $S$,
the time $t$, the metric $a_t$, the 1-form $b_t$ and the potential
$\mathcal{U}$, so as to express and study Eisenhart's theorem. We
shall see that natural coordinates do not exist and that, in order
to determine $b_t$, one has to fix the Newtonian flow (frame) on
$E$. Indeed, the 1-form $b_t$ will play the role  of Coriolis
potential, determining the rotational inertial forces of the
dynamical system. The dependence of $b_t$ on the frame is then
expected as the inertial forces depend on how the frame chosen moves
on the classical spacetime $E$. We shall also prove that the flow
over $E$ can be equivalently replaced by an Abelian connection on a
suitable lightlike principal bundle, its curvature giving the
fictitious inertial forces associated to the Newtonian frame.

The Eisenhart metric takes its simplest and most symmetric form in
the case of a free particle in Euclidean space $a_{bc}=\delta_{bc}$,
$b_c=0$, $\mathcal{U}=const$. Remarkably, in this case the Eisenhart
metric becomes the Minkowski metric.
While it will be included in our analysis, the flat case is most
interestingly studied by using a group theoretical approach
\cite{minguzzi05e}. In \cite{minguzzi05e} it has been argued that
ordinary shadows behave very much like classical objects living in a
quotient classical spacetime thus obeying a Galilean invariant
physics. Related differential geometric studies of shadows had
previously appeared
 in \cite{jordan61} and \cite[Th. 4.1]{sachs61} (see also the account \cite[Sect. 5.3.2]{frolov98}).

The paper is organized as follows. In section \ref{die} the
spacetime geometry induced by the presence of  a lightlike Killing
field is studied in detail. The classical spacetime is identified
with the corresponding quotient manifold. It is in this section that
most of the classical physical concepts such as absolute
simultaneity, absolute time, space metric, Newtonian frame and
Coriolis inertial forces are introduced. In section \ref{ceis} the
light lift is introduced, its relation with the classical action
explained and the causal version of Eisenhart's theorem is stated.
In the short section \ref{sfg} the Eisenhart's theorem is proved. In
section \ref{ceis3} the Eisenhart's theorem in the lightlike case is
related to a kind of Fermat's principle in the case of  lightlike
target curves, and the existence of minimizers for the classical
action is shown to follow from the causal simplicity of Eisenhart's
spacetime. Finally, in section \ref{conc} some conclusions are drawn
and in the appendix  a brief introduction to Newton-Cartan theory is
given so as to clarify the differences in aim an motivations between
this work and those on that theory.

Our notations are as follows. Let $d \in \mathbb{N}$. The indices
$i,j,k$, take the values $1,\ldots, d+1$,  the indices $a,b,c$, take
the values $1,\ldots, d$.  The Greek indices
 $\alpha,\beta,\mu,\nu$,
take the values $0,1,\ldots, d+1$, and the indices ${ A},{ B},{C}$,
take the values $0,1,\ldots, d$. We use the spacelike convention
$\eta_{00}=-1$, and units such that $c=1$. The exterior product of
two 1-forms $\alpha$, $\beta$, is $\alpha \wedge \beta=\alpha
\otimes \beta- \beta \otimes \alpha$. The generic point of the
Lorentzian manifold $M$ is usually denoted with $x$ or $m$, that of
the extended configuration space $E$ with $e$, and that of the
configuration space $S$ with $q$. By {\em lightlike}, {\em causal}
or {\em timelike} curve we always mean a regular curve. For basic
results in Lorentzian geometry needed in this work and especially in
the proof of theorem \ref{cas} the reader is referred to
\cite{minguzzi06c,beem96,hawking73}.

\section{Lightlike Killing vector fields} \label{die}

In this section we study the geometry induced on spacetime by the
presence of a lightlike Killing vector field (for related
investigations see \cite{lukacs81,zeghib98}). As we shall see a
lightlike Killing field allows one to construct a quotient manifold
$E$ in which the physics is Newtonian provided the physics on the
original manifold $M$ was Einstenian. We shall not go into all the
details which stay at the heart of this correspondence between
general relativity and the so called Newton-Cartan theory on the
quotient spacetime \cite{kunzle72}. The reader is referred to
\cite{duval85,kunzle86,duval91,bernal03b,julia95} for more on this
subject.  As a contribution, using coordinate independent methods,
we shall prove the equivalence between Newtonian flows on the
quotient manifold and Abelian connections.

Consider a  $(d+1)+1$-dimensional Lorentzian spacetime\footnote{The
manifold is at least $C^k$, and the metric is at least $C^{k-1}$,
$k=3$, but $k=2$ will suffice almost everywhere unless in those
parts where the existence and continuity of the Riemann tensor is
invoked. The tensor fields $a_t$, $b_t$, $U$ and $\psi$ (to be
introduced later) have the same degree of differentiability of the
metric.} $(M,g)$ with a future directed lightlike Killing vector
field $n$, $n_{\mu;\nu}+n_{\nu;\mu}=0$. The previous equation
implies that the field $n$ is geodesic $n_{\nu;\mu}n^{\mu}=0$,
divergenceless $ n^{\mu}_{; \mu}=0$ and `shear free' by definition.
By \cite[Lemma 3.2]{beem88}, $n$ is nowhere vanishing. We introduce
an equivalence relation  between events that lie on the same
geodesic integral line, and denote with $E$ the quotient space $\pi:
M \to E$. The set $E$ will also be called the {\em extended
configuration space}, or the {\em classical spacetime}.

Now we have to add some conditions to make $E$ a $(d+1)$-dimensional
manifold and $\pi$ a differentiable projection. Unfortunately, even
if $n$ is a complete vector field and $M$ is a causal spacetime $E$
may not be a manifold. Indeed, there are pathological examples in
which the geodesic integral lines of $n$ come arbitrarily close to
themselves, preventing $E$ from being a manifold. An example can be
constructed from  2+1 Minkowski spacetime $(\Lambda,\eta)$,
$\eta=\dd q^2-\dd t \otimes \dd y- \dd y \otimes \dd t -\dd t^2$ by
making the identifications $(t,q,y) \sim (t,q+1,y)$, and $(t,q,y)
\sim (t,q+\alpha,y+1)$ where $\alpha$ is an irrational number.

One could try to impose some stronger causality constraints on
$(M,g)$ such as the distinguishing or strong causality properties,
however they would be too restrictive for later applications.

Thus we simply assume that $E$ is a manifold, that $\pi$ is
differentiable and that $M$ has locally a direct product structure,
i.e. $ \mathcal{A}\times \mathbb{R}$, $\mathcal{A}$ open set on $E$.
In other words we assume that $M$ is a principal bundle with
structure group $(\mathbb{R},+)$ and fiber diffeomorphic to
$\mathbb{R}$. In particular $n$ is a complete vector field. As the
fiber is contractible, global sections exist \cite{nash82} and hence
this local assumption implies that $M$ is diffeomorphic to $ E\times
\mathbb{R} $. We shall refer to these properties by saying that
$(M,g)$ has a principal bundle structure.

By $n$ we denote, depending on the context, the Killing vector field
or the 1-form field obtained by lowering an index with $g$, $n_{\mu}
\dd x^{\mu}$.

Our next assumption is that the 1-form field $n$ determines a
distribution of null hyperplanes, ${\rm ker}\, n$, which is locally
integrable, that is
\begin{equation} \label{ho}
n \wedge \dd n=0.
\end{equation}
These requirements are weaker that those needed to define a Bargmann
structure \cite{duval85,duval91}, since there $n$ must be
covariantly constant,  $n_{\mu;\nu}=0$, and hence closed $\dd n=0$.

The following theorem, which clarifies earlier results
\cite{ehlers62,julia95} (see also \cite{ihring75}), relates the
hypersurface orthogonality condition (\ref{ho}) with an Einstein
equation on $M$.  \\

\begin{theorem}Let $n$ be a lightlike Killing vector field on the spacetime
$(M,g)$. In any spacetime dimension $n \wedge \dd n=0$ if and only
if $R_{\mu \nu}n^{\mu} n^{\nu}=0$. In particular if the spacetime
has dimension lower than $3$ then $n \wedge \dd n=0$, and $R_{\mu
\nu}n^{\mu} n^{\nu}=0$ whereas if $M$ is 4-dimensional ($d=2$), then
$R_{\mu \nu}n^{\mu} n^{\nu} \ge 0$ and
\begin{equation} \label{bhg}
z_{\alpha}:=2\epsilon_{\alpha \beta \gamma \delta} n^{\beta}
n^{[\gamma;\delta]}=\pm (8 R_{\mu \nu}n^{\mu} n^{\nu})^{1/2} \,
n_{\alpha} ,
\end{equation}
where the sign is constant and well defined in those disconnected
open sets where $R_{\mu \nu}n^{\mu} n^{\nu} > 0$.
\end{theorem}

\begin{proof}
From $n_{\beta;\alpha;\mu}-n_{\beta;\mu;\alpha}=R_{\beta \nu \mu
\alpha} n^{\nu} $, and permuting repeatedly the indices and summing
and subtracting the equations so obtained we have (Killing vector
lemma) $n_{\beta;\alpha;\mu}=-R_{\mu \nu \alpha \beta} n^\nu$. Thus
\begin{eqnarray}
n^{\alpha;\beta}n_{\alpha;\gamma}=(n^{\alpha}n_{\alpha;\gamma})^{;\beta}-n^{\alpha}n_{\alpha;\gamma}^{\
\ \ \, ;\beta}=R^{\beta}_{\ \nu \gamma \alpha } n^{\nu} n^{\alpha} ,
\end{eqnarray}
In particular $n_{[\alpha;\mu]} n^{[\alpha;\mu]}=R_{\mu \nu}n^{\mu}
n^{\nu}$. Note that $n^{\mu} n_{\mu}=0=n_{[\alpha;\mu]}n^{\mu}$.

Let
\[ w_{\gamma \mu  \nu}=n_{\gamma} n_{[\mu ; \nu]}+n_{\nu} n_{[\gamma ;
\mu]}+n_{\mu} n_{[\nu ; \gamma]} ,
\]
so that $n \wedge \dd n=0$ iff $w_{\gamma \mu  \nu}=0$. The tensor
$w_{\gamma \mu  \nu}$ is skew-symmetric.  Note that $w_{\gamma \mu
\nu} n^{[\mu ; \nu]} = n_{\gamma}(R_{\mu \nu}n^{\mu} n^{\nu})$, thus
$n \wedge \dd n=0$ implies $R_{\mu \nu}n^{\mu} n^{\nu}=0$. For the
converse, let $R_{\mu \nu}n^{\mu} n^{\nu}=0$, note that $w_{\alpha
\mu \nu} w^{\beta \mu \nu}=n_{\alpha} n^{\beta} (R_{\mu \nu}n^{\mu}
n^{\nu})=0$. Let $\{ x^{\mu} \}$ be coordinates at the chosen
arbitrary event $x$ (or introduce a non-holonomic orthonormal frame)
so that $g_{\mu \nu}=\eta_{\mu \nu}$ at that event. Then $0=w_{0 \mu
\nu}w^{0 \mu \nu} =-\sum_{\mu, \nu \ne 0} (w_{0 \mu \nu})^2$ hence
$w_{0 \mu \nu}=0$. The equation $w_{\alpha \mu \nu} w^{\alpha \mu
\nu}=0$, now reads $\sum_{\alpha,\mu, \nu \ne 0} (w_{\alpha \mu
\nu})^2=0$ and thus all the components of $w_{\alpha \mu \nu}$
vanish, i.e. $n \wedge \dd n=0$.

Let us consider some particular cases. If the dimension of the
spacetime is lower or equal to 2 then  $n \wedge \dd n=0$ follows
for dimensional reasons. If the spacetime has dimension 3 defined
$z=\epsilon_{\beta \gamma \delta} n^{\beta} n^{[\gamma;\delta]}$, we
have $z^2=-\delta_{\eta \alpha \beta }^{\delta \mu \nu}
 \, n^{\eta}\, n_{\delta}\, n^{[\alpha;\beta]} \,n_{[\mu;\nu]} =0 $.

If the spacetime has dimensions 4 then the following quantity
vanishes
\begin{equation}
z_{\sigma}z^{\sigma}=-4 \delta_{\eta \alpha \beta }^{\delta \mu \nu}
 \, n^{\eta}\, n_{\delta}\, n^{[\alpha;\beta]} \,n_{[\mu;\nu]} =0 .
\end{equation}
Since $z^{\alpha}$ is a lightlike vector and $z_{\alpha}
n^{\alpha}=0$, it follows that $z^{\alpha}$ must be proportional to
$n^{\mu}$, $z^{\alpha}=f n^{\alpha}$. The identity
\begin{eqnarray}
f^2 n_{\mu}&=&fz_{\mu}=2 \epsilon_{\mu \nu \alpha \beta} z^{\nu}
n^{[\alpha;\beta]} =4 \delta_{\mu \alpha \beta}^{\gamma \eta \sigma}
n_{\gamma} \, n_{[\eta;\sigma]} \, n^{[\alpha;\beta]} \nonumber \\
&=&8 n_{\mu} (n_{[\alpha;\beta]}n^{[\alpha;\beta]})=(8 R_{\alpha
\beta}n^{\alpha} n^{\beta}) n_{\mu} ,
\end{eqnarray}
shows that $R_{\alpha \beta}n^{\alpha} n^{\beta}\ge 0$ and $f=\pm (8
R_{\mu \nu}n^{\mu} n^{\nu})^{1/2}$ where, by continuity, the sign
$\pm$ is constant in those regions where $R_{\mu \nu}n^{\mu} n^{\nu}
>0$.

\end{proof}

This result shows that the single Einstein's vacuum equation $R_{\mu
\nu}n^{\mu} n^{\nu}=0$ in the  spacetime $(M,g)$ is equivalent to
the hypersurface `orthogonality' of the lightlike Killing vector
field.

The problem of reducing the Einstein equations to $E$ has been
studied in generality in \cite{julia95}. Some other results will be
mentioned in what follows. We do not impose the full Einstein's
equations on $M$ because they would lead to a quite restrictive
geometry on $E$.

\subsection{Absolute simultaneity, absolute time, Newtonian frames
and space metric} \label{die2}

Due to Eq. (\ref{ho}), by Frobenius theorem there are local
functions $\psi>0$ and $t$ such that $n=-\psi \dd t$. In principle
the construction that follows could be developed locally, but in
order to simplify the analysis and the notation we shall assume that
the distribution of hyperplanes $\textrm{ker} \,n$, is not only
locally integrable, but also globally integrable, that is, there are
global functions $\psi>0$ and $t$ such that $n=-\psi \dd t$. If $M$
is simply connected this condition is automatically satisfied by a
Bargmann structure (as $\dd n=0$).

It is time to pause and to recall all the assumptions that will not
be reminded in what follows. $M$ is a Lorentzian time oriented
manifold (i.e. a spacetime) endowed with a complete future directed
lightlike Killing vector field $n$, which makes $(M,g)$ a principal
bundle of base $E$, and whose associated distribution of hyperplanes
$\textrm{ker} \,n$ is globally integrable.

\begin{figure}
\centering \psfrag{O}{$\omega_t$} \psfrag{X}{$\!\!\!\! \!\{q^a\}$}
\psfrag{P}{$\pi$} \psfrag{XD}{$y$} \psfrag{T}{$t$} \psfrag{M}{$M$}
\psfrag{N}{$n$} \psfrag{S}{$\!\!S$} \psfrag{SI}{$\Sigma_t$}
\psfrag{SIG}{$\Sigma$} \psfrag{Q}{$\!E$} \psfrag{NT}{$N_{t}$}
\psfrag{ST}{$S_t$}
\includegraphics[width=6cm]{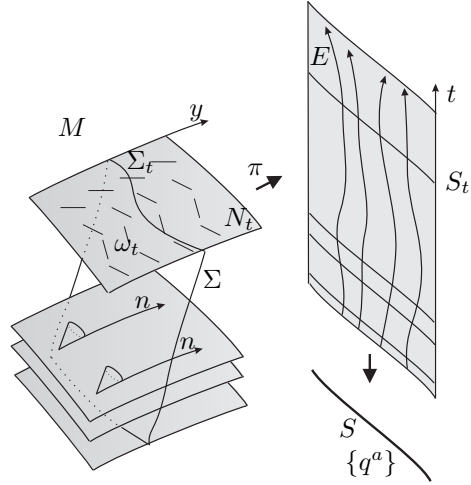}
\caption{The spacetime $(M,g)$ due to the condition $n \wedge \dd
n=0$ on the lightlike Killing field $n$ splits into principal
bundles $\pi_t : N_t \to S_t$. A $t$-parameter family of connection
1-forms $\omega_t$ on those bundles (displayed here with their
kernel on the tangent spaces $TN_t$) is equivalent to a Newtonian
extended frame on $E$. The connections $\omega_t$ have in general an
holonomy which, unfortunately, can not be displayed in this figure
since $S$ here is one-dimensional.} \label{figure1}
\end{figure}

The function $t$ is such that  given  a future directed timelike
vector $V$, it is $\dd t[V]=-g(n,V)/\psi>0$, because $n$ and $V$ are
in the same forward light cone. Thus $t$ strictly increases over
timelike curves, but it is constant on the null geodesics generated
by $n$ and in particular the submanifolds $t(x)=\bar{t}$ are null
hypersurfaces. The function $t$ satisfies the requirements for a
{\em quasi-time function} in the sense of \cite[Def. 13.4]{beem96},
nevertheless we shall call it simply {\em time function} not because
of its role on $M$ but because of the role that it will play on $E$
(see figure \ref{figure1}). Indeed, $\dd t[n]=\p_{n}t=0$ so that $t$
is a function on $E$ lifted to $M$. We shall denote with the same
letter $t$ the time function on $E$ and its lift to $M$. The
condition $L_{n} n =0$ implies $\p_{n}\psi=0$, thus $\psi$ depends
only on the base point too.

The global integrability condition implies the existence of a
natural {\em absolute simultaneity} on $E$, determined by the
surfaces $t=cnst.$ However, note that the time function is not, by
itself, completely determined since one can change time function $t$
to $t'=f(t)$, $f'>0$, by suitably changing $\psi: E \to \mathbb{R}$.
For this reason, in what follows, we shall always keep the field
$\psi$ aside a time differential, for instance in expressions like
$\psi \dd t$ or (when it will make sense) $\frac{1}{\psi}
\frac{\p}{\p t}$.

Each null hypersurface $N_{\bar{t}}$ on $M$, of equation
$t(x)=\bar{t}$, may be regarded as a principal bundle over
$S_{\bar{t}}$ (i.e. the locus $t(x)=\bar{t}$ on $E$) with fiber
$\mathbb{R}$ and structure group $(\mathbb{R},+)$ generated by $n$.

If we are in a Bargmann structure, $\dd n=0$, then $\psi$ depends
solely on $t$, and we can choose the new time parameter $t'$, $\dd
t'=\psi \dd t$, so that $\psi \to 1$. We conclude that a Bargmann
structure has a natural {\em absolute time} on $E$, i.e. the one
that makes $\psi=1$.

Let $m \in M$ and $e=\pi(m) \in E$. The projection $\pi$ is a
submersion, that is for any tangent vector $v$ at $e$ there is a
tangent vector $V$ at $m$ that projects on it, $v=\pi_{*}V$.
Moreover, $n$ spans the kernel (vertical space) of this projection.
Any tangent vector $v$ at $e$ can therefore be represented in the
full spacetime with the equivalence class $[V]$, where $V \in
TM_{m}$ is a representative and any alternative representative reads
$V'=V+\alpha n$ for a suitable $\alpha \in \mathbb{R}$.

In order to construct coordinates on $M$ we first need natural
coordinates on $E$.

\begin{definition}
By {\em  Newtonian extended reference frame}, or simply {\em
Newtonian frame}, we mean a  future directed  (at least $C^1$)
vector field $w(e)$ on $E$ (future directed means $\dd t[w]>0$)
normalized in such a way that $\psi \dd t[w]=1$.
\end{definition}

In $M$ the Newtonian  frame is represented by a field of equivalence
classes $[W(m)]$ where $W(m)$ is a representative vector field such
that $g(n,W)=-1<0$ (any representative field satisfies this
relation).
%

The integral lines of $w(e)$ on $E$ form a $d$-dimensional quotient
space $S$ that we call the {\em space} of the reference frame. We
cover $S$ with charts of coordinates denoted $q^{a}$, and $(t,q^a)$
become coordinates on $E$. We stress that natural coordinates (up to
transformations ${q'}^a={q'}^a(q^b)$) on $E$ are given only if a
choice of Newtonian frame has been made. In these coordinates the
normalized vector field that defines the frame is
$w=\frac{1}{\psi}\frac{\p}{\p t}$ (one should be careful because
this expression makes sense on $E$ but it does not make sense on $M$
since the last coordinate $y$ has not yet been defined). The
Newtonian frame gives another fibration $\pi_{ES}: E \to S$, and the
quotient space $S$ can also be identified with any chosen slice
$S_t$, $t=cnst.$ on $E$. We have $E=T\times S$ where $T$ is the
image of the function $t: E \to \mathbb{R}$. It is an open connected
subset of $\mathbb{R}$ which can be different from $\mathbb{R}$, as
it also depends on the choice of $\psi$. In the case of a Bargmann
structure, as $t$ is uniquely determined, it is natural to ask that
$T=\mathbb{R}$.

The metric $g$ induces a (Newtonian frame independent) Riemannian
metric on $S_t$ as follows. Let $w, v$ be tangent vectors to $S_t$
at $q \in S_t$, then  define the Riemannian metric
\begin{equation} \label{aaa}
a_t(w,v)=g(W,V) ,
\end{equation}
where $W$ and $V$ are any representatives on $N_t$. The right hand
side is independent of the representatives, since as $w$ and $v$ are
tangent to $S_t$ we have $V^{\mu}n_{\mu}=W^{\mu} n_{\mu}=0$, i.e.
$V$ and $W$ are tangent to $N_t$. The condition $a_t(v,v)=0$ implies
$g(V,V)=0 \Rightarrow V \propto n \Rightarrow v=0$, that is $a_t$ is
positive definite. In a coordinate chart $(t,q^a)$ adapted to the
Newtonian frame the metric on $S_t$ may be written
\begin{equation}
a_t(w,v)=a_{ab}(t,q)\dd q^a \dd q^b .
\end{equation}
Note that while the {\em space} $S$ is a Newtonian frame dependent
concept, the {\em spaces} $S_t$ endowed with the metrics $a_t$ are
not. Due to the identification of the different $S_t$ on a single
space $S$  provided by the Newtonian reference frame, the metric
$a_t$ can also be regarded as a time dependent metric on $S$.

\subsection{Equivalence between Newtonian frames and Abelian
connections} \label{die4}

  Interestingly, the Newtonian frame $v(e)$ determines a connection
(in fact a one-parameter family, one for each choice of $t$) over
the principal bundle $\pi_t: N_{t} \to S_t$
($\pi_t:=\pi\vert_{N_t}$). Indeed, let $V_t$ be a representative
vector field restricted to $N_{t}$, $V_t=V\vert_{N_{t}}$, and
construct the 1-form $\omega_{t}(\cdot)=- g(\cdot,V_t)\vert_{N_t}$
which acts on the vectors tangent to $N_t$. It is easy to see that
if $W$ is tangent to $N_{t}$ at $m$, $n_{\mu}W^{\mu}=0$, then
$\omega_{t}(W)$ is independent of the representative $V_t$ chosen,
and depends only on the Newtonian frame at time $t$. Moreover, the
following properties can be easily checked taking into account
that, by construction, $L_n V \propto n$,
\begin{itemize}
\item[(a)] $\omega_t(n)=1$, \\
\item[(b)] $L_{n}\omega_t=0$.
\end{itemize}
Thus, $\omega_t$ is a connection 1-form on $\pi_t: N_{t} \to S_t$,
\cite{kobayashi63}. One should be careful because $\omega_t$ is not
 defined on vectors which are not tangent to $N_t$.

Conversely, any ($C^1$) one-parameter family of connections
$\omega_t$ on the principal bundles $\pi_t: N_{t} \to S_t$ is
associated to a Newtonian frame on $E$. Since the proof is somewhat
more involved we give it in the form of a theorem.

\begin{theorem} Let $v(e)$ be a future directed vector field on $E$
such that $\psi \dd t[v]=1$ (Newtonian frame), and let $V(m)$ any
vector field on $(M,g)$ which projects on it, then
$\omega_{t}(\cdot)=- g(\cdot,V)\vert_{N_t}$ (a 1-form which acts on
$TN_t$) is a connection for the principal bundle $\pi_t: N_t \to
S_t$ which is independent of the representative $V(m)$.

Conversely, let $\omega_t$ be a ($C^1$) one-parameter family of
connections for the principal bundles $\pi_t: N_t \to S_t$ (i.e.
1-form fields over $N_t$ which satisfy (a) and (b) above). There is
a unique Newtonian frame $v(q)$ on $E$ such that for any
representative field $V$, $\pi_{*} V= v$, it is
$\omega_t(\cdot)=-g(\cdot, V)\vert_{N_t}$.

\end{theorem}

\begin{proof}
The first part has been already proved. Introduce an arbitrary
vector field $W$ on $M$ such that $-g(n,W)>0$.
%
Extend $\omega_t$ to a 1-form field  $\omega$ over $M$ such that,
(i) $\omega_t(\bar{W})=\omega(\bar{W})\vert_{N_t}$,  for every
$\bar{W}$ such that $n_{\mu}{\bar{W}}^{\mu}=0$, i.e. the restriction
to the manifold $N_t$ coincides with $\omega_t$ and, (ii)
$\omega(W)=0$. Since $g$ is invertible there is a unique vector
field $V$ such that $\omega(\cdot)=-g(\cdot, V)$. The field $V$ is
not tangent, at any point, to $N_t$, indeed
$n_{\mu}V^{\mu}=-\omega[n]=-\omega_t[n]=-1 \ne 0$. Moreover,
condition (b) implies that $L_n V=0$, hence $V$ is projectable
 on a vector field $\pi_{*}V=v$ on $E$ such that $\psi\dd t[v]=1$.
Note that by changing $V \to V+\alpha(m) n$, the 1-form $-g(\cdot,
V)$ restricted to the manifold $N_t$ still coincides with
$\omega_t$. Now, let us show that the choice of $W$ does not change
the field $v$. The reason is that if $W \to W'$, then $V \to V'$
where
\begin{equation}
V'=V-\frac{g(V,W')}{g(n,W')} n .
\end{equation}
We conclude that the choice of $W$ which serves to extend $\omega_t$
fixes only the vertical part of $V$ and does not affect its
projection.
\end{proof}

In summary, we proved that there is a one-to-one correspondence
between Newtonian frames on $E$ and  one-parameter families of
connections $\omega_t$ on the principal bundles $\pi_t: N_t \to
S_t$.

Note that if a Newtonian frame on $E$ reads $w=\frac{1}{\psi}
\frac{\p}{ \p t}$, then one of its representative $W$ on $M$, takes
the same form $W=\frac{1}{\psi}\frac{\p}{ \p t}$ irrespective on how
the coordinate $y$ is defined on $M$ (provided $n=\p/\p y$).

Each manifold $N_t$ has a natural exterior differential $\dd_t$. The
distinction between $\dd_t$ and $\dd$ defined over $M$  must be kept
in mind. In short $\dd_t$ does not differentiate with respect to the
$t$ function. For instance $f(t)$ has a non-vanishing exterior
differential $\dd f\ne 0$ but $\dd_t f=0$, because on the manifold
$N_t$, the function $f$ is constant.

We can now define the curvature $\Omega_t=\dd_t \omega_t$ on each
principal bundle $\pi_t: N_t \to S_t$. The proved characterization
of Newtonian frames allows us to define the {\em non-rotating}
Newtonian frames as those associated with a flat connection,
$\Omega_t=0$. Indeed, the curvature $\Omega_t$ will be strictly
linked to the Coriolis inertial forces.

\begin{table}
    \begin{center}
    \setlength\extrarowheight{3pt}
        \begin{tabular}{|p{5cm}|p{5cm}|}
            \hline
             & \\
               $\qquad \ \ \, $ {\bf Gauge theory}& $\ \ \ $ {\bf Physical interpretation}  \\
             & \\
            \hline\hline
          $ \ $ \small{Lightlike Killing field $n$ on $(M,g)$. } \small{Principal bundle $\pi: M \to E$ }  & $ \ $  \small{Existence of a quotient space}  \small{$E$ (the Newtonian spacetime)}\\
                                   \hline
           $ \ $   \small{$n \wedge \dd n=0$. One-parameter }  \small{family  of principal bundles }   \small{$\pi_t:N_t \to S_t$ } & $ \ $   \small{Absolute simultaneity on $E$. } \small{A particular vacuum  Einstein} \small{equation on $M$ is satisfied}   \\
                                    \hline
            $ \ $  \small{Bargmann structure, $\dd n=0$}& $ \ $ \small{Absolute time on
            $E$}
             \\
            \hline
            $ \ $  \small{Connection $\omega_t$}  & $ \ $  \small{Newtonian frame on $E$}\\
                         \hline
           $ \ $  \small{Curvature $\Omega_t$} &  $ \ $ \small{Coriolis inertial forces }   \\
            \hline
            $ \ $  \small{ $\Omega_t=0$, flat bundle.} &  $ \ $ \small{Non-rotating Newtonian frame. }   \\
            \hline
                                \end{tabular}
    \end{center}
    \label{table}
\end{table}

\subsection{The Coriolis potential and the Eisenhart metric} \label{die5}

Consider   the image $\Sigma$, of a ($C^1$) section $\sigma: E \to
M$ (global sections exist since the fiber is contractible
\cite{nash82}), and define the function $y: M \to \mathbb{R}$, as
the unique function such that $y=0$ over $\Sigma$, and $\dd y[n]=1$.
Because $n$ is complete, the image of $y$ is all $\mathbb{R}$ and
hence $M=E\times \mathbb{R}=T\times S\times\mathbb{R} $.

Let $\pi_S=\pi_{ES} \circ \pi$, $\pi_S: M \to S$. Given a section
$\sigma$, and a coordinate chart $\{q^a\}$ defined on an open set
${B}\subset S$, the natural coordinates on $\pi_S^{-1}B \subset M$,
are then $\{x^{\mu}\}=\{t,q^1,\ldots,q^{d},y\}$.

The section $\sigma$ can also be regarded as a 1-parameter family of
sections $\sigma_t$ of the bundles $\pi_t: N_t \to S_t$, or, using
the diffeomorphism between $S$ and $S_t$ provided by the Newtonian
frame, as a 1-parameter family of sections $\sigma_t: S \to N_t$ of
the 1-parameter family of principal bundles $\pi_t: N_t \to S$.

Thus over $S$ we can define not only the time dependent metric
$a_t$, but also the time dependent 1-form field
\begin{equation}\label{bbb}
b_t=-\sigma^{*}_t\omega_t ,
\end{equation}
which is the (minus) potential of the connection $\omega_t$ on the
principal bundle $\pi_t: N_t \to S$. If  $a_t$ and $b_t$ are known
to be independent of $t$ we shall omit the index $t$. The
corresponding curvature reads in coordinates $\dd_t b=(\p_{[a}
b_{b]})\dd q^a \wedge \dd q^b=-\sigma_t^{*}\Omega_t$.

We claim that there is a time dependent function $U:E \to
\mathbb{R}$,  such that the metric $g$ on $M=E\times
\mathbb{R}=T\times S\times\mathbb{R} $ reads (recall that
$n=-\psi\dd t$)
\begin{equation} \label{asd}
g=\pi_S^{*}a_t+ n \otimes (\dd y-\pi_S^{*}b_t)+(\dd y-\pi_S^{*} b_t)
\otimes n -(2U+1) n^2 ,
\end{equation}
where $\pi_S^{*}$ is the pullback of covariant tensors from $S$ to
$M$, and where for simplicity the pullback of functions from  $E$ to
$M$ does not receive a different symbol (thus, as done with
functions $\psi$ and $t$, we replaced $\pi^{*}U$ with $U$). Note
that $\omega_t$ is undefined over vectors which are not tangent to
$N_t$, on the contrary $\pi_S^{*}b_t$ is perfectly meaningful. This
metric reduces to that considered by Eisenhart in the case of a
Bargmann structure. The generalized form with $\psi$ not constant
was introduced in \cite{lichnerowicz55}, again as a pure coordinate
dependent definition.

The particular form $-(2U+1)$ for the coefficient in front of
$(-\psi \dd t)^2$ is chosen in order to obtain a nice Minkowskian
limit for $U=0$, $S=\mathbb{E}^{3}$, $a_{t\, ab}=\delta_{ab}$,
$b=0$, $\psi=1$, with the usual shadow coordinates (see
\cite{minguzzi05e}).

In order to prove Eq. (\ref{asd}) let us first consider two vectors
$V$ and $W$ at $m \in M$, tangent to $N_t$ so that only the first
term in the right-hand side of Eq. (\ref{asd}) does not vanish. We
have (due to the identification between $S$ and $S_t$, $\pi_{S*}$
becomes $\pi_{*}$)
\begin{equation}
\pi_S^{*}a_t(V,W)=a_t(\pi_*V,\pi_*W)=g(V,W) ,
\end{equation}
where in the last step we used the fact that $V$ is a representative
for $\pi_*V$, and analogously for $W$. Thus the right-hand side of
Eq. (\ref{asd}) restricted to the vectors tangent to $N_t$ coincides
indeed with the metric $g$.

We are now going to show that there is a vector field $W$ such that
the scalar product of $W$ with a vector tangent to $N_t$ made with
the right-hand side of Eq. (\ref{asd}) coincides with that made with
the metric. Let $w$ be the Newtonian frame field that defines $S$,
$\psi \dd t[w]=1$, and let $W$ be the unique representative,
$\pi_{*}W=w$, such that $\dd y[W]=0$ (it can be obtained by Lie
transporting $\sigma_{*}w$ over the fibers). Clearly, $\pi_{S*}W=0$
and $n[W]=-1$. Thus if $V$ is tangent to $N_t$ the scalar product
made with the right-hand side of Eq. (\ref{asd}) reads
\begin{equation}
-(\dd y-\pi_S^{*}b_t)[V]=-(\dd
y[V]-b_t[\pi_{*}V])=-\omega_t[V]=g(W,V).
\end{equation}
The function $U$ is then chosen such that $g(W,W)=-(2U+1)$ which
proves the claim. The proof also shows that $(\dd y-\pi_S^{*}b_t)$
is an extension of the connection 1-form $\omega_t$, in the sense
that $\omega_t$ can be regarded as the restriction of $(\dd
y-\pi_S^{*}b_t)$ to the vectors tangent to $N_t$.

Given the definition
\begin{definition} \label{cft}
Let $S$ be a $d$-dimensional manifold endowed with a (all possibly
time dependent and at least $C^1$) positive definite metric $a_t$,
1-form field $b_t$ and scalar fields $U$ and  $\psi>0$. Let $T$ be a
connected open subset of $\mathbb{R}$ and set $M=T \times S \times
\mathbb{R}$, where $x=(t,q,y)$ is the generic event of $M$ and
define $n=-\psi\dd t$  on $M$. The Eisenhart spacetime is the
manifold $M$ endowed with the metric (\ref{asd}) and time
orientation such that $n^{\mu}\p/\p x^{\mu}$ is future directed.
\end{definition}

\noindent the results obtained can be expressed, somewhat
synthetically, as follows

\begin{theorem}
Let $(M,g)$ be a $(d+1)+1$-dimensional spacetime endowed with a
complete lightlike Killing vector field $n$ which makes $(M,g)$ a
$(\mathbb{R},+)$ principal bundle and such that the distribution of
hyperplanes, $\textrm{ker}\, n$, is globally integrable. Let $N_t$
be the corresponding foliation with $n=-\psi \dd t$, $\psi>0$, $T$
the image of $t$, and let $S_t$ be the quotient spaces under the
action of $n$, so that $\pi_t: N_t \to S_t$ are principal
$(\mathbb{R},+)$-bundles. Let $\omega_t$ be a ($C^1$) 1-parameter
family of connections on the bundles and let $S$ be the space
obtained identifying the different $S_t$ through the action of the
Newtonian frame associated to $\omega_t$. Over $S$ a time dependent
space metric $a_t$, can be defined through Eq. (\ref{aaa}) and,
given a ($C^1$) 1-parameter family of sections $\sigma_t: S_t \to
N_t$, a time dependent 1-form field $b_t$ can be defined through Eq.
(\ref{bbb}). Then $M=T\times S\times \mathbb{R}$ and a function $U$
exists such that the metric $g$ takes the form (\ref{asd}). The
spacetime $(M,g)$ is then a Eisenhart spacetime.
\end{theorem}

A change of section (gauge transformation) $\sigma \to \sigma'$,
($\sigma_t \to \sigma'_t$) reads
\begin{eqnarray}
y'&=& y+\alpha(t,q), \label{pz1} \\
b'_t&=&b_t+\dd_t \alpha , \label{pz2} \\
\psi U'&=&\psi U-\p_t\alpha, \label{pz3}
\end{eqnarray}
and $\omega_t$ is invariant as expected. The gauge transformations
do not change the Newtonian frame but only the way in which the
events of $M$ are labeled.

A special class of gauge transformations are those for which
$\alpha=Ct$ with $C$ constant. Indeed, they leave the gauge
potential $b_t$ unchanged but change by an additive constant $\psi
U$, and hence any function $\mathcal{U}$ of the form
$\mathcal{U}=\psi U+K_1 \psi+K_2$, $K_1,K_2 \in \mathbb{R}$. Indeed
we shall see that suitable constants $K_1$ and $K_2$ exist which
allow us to interpret $\mathcal{U}$ (and hence $U$ for a Bargmann
structure) as the classical potential. However, one should keep in
mind that $U$ (and hence $\mathcal{U}$) is not uniquely determined
because of the existence of the wider class of gauge transformations
given above. Nevertheless, if the connections $\omega_t$ are flat a
section $\sigma$ exists such that, for every $t$, $b_t=0$. For this
special class of non rotating Newtonian frames it is possible to
identify uniquely (up to an additive time function) a {\em potential
function} $\mathcal{U}$.

A trivial consequence of definition \ref{cft} is

\begin{lemma} \label{xcv}
The Eisenhart spacetime is a causal spacetime.
\end{lemma}

\begin{proof}
Let $V(\lambda)=\p/\p \lambda$ be the tangent vector to a future
directed causal curve $x(\lambda)$. Since $V$ and $n$ are future
directed $g(n,V)\le 0$ or $\dd t[V]\ge 0$ where the equality sign
holds iff $V \propto n$. Thus as $t(\lambda)$ is a non-decreasing
function the causal curve can not be closed unless
$t(\lambda)=cnst.$ in which case $x$ is a geodesic generated by $n$
which is not closed neither.
\end{proof}

%

\subsection{The  $\frac{1}{\mu}\,$-lift.}

Given a future directed field $v(e)$ on $E$ the representative field
$V$ on $M$, $\pi_{*}V=v$, is not uniquely defined. Nevertheless, it
can be uniquely identified by imposing some conditions on its norm.
We have the following

\begin{lemma}
Let $V(m)$ be a representative field on $M$ which projects on a
future directed field $v(e)$ on  $E$, and let
$\omega(\cdot)=-g(\cdot,V)$. The following conditions are equivalent
\begin{itemize}
\item[(i)] $L_n V=0$,
\item[(ii)] $L_n \omega=0$,
\item[(iii)] $L_n g(V,V)=0$,
\end{itemize}
and $V(m)$ is uniquely identified if they hold and the function
$g(V,V)(e)$ on $E$ is given. In particular $V(m)$ can be chosen to
be causal, and for any given function $\frac{1}{\mu}: E \to
[0,+\infty)$,  it is uniquely identified by imposing
\begin{equation} \label{lift3}
\frac{1}{\mu}=\frac{\sqrt{-g(V,V)}}{-g(n,V)} ,
\end{equation}
in which case $V$ is said to be the $\frac{1}{\mu}$-lift of $v$.
\end{lemma}

\begin{proof}
Since $V$ is a representative, $L_n V=\beta n$ for a suitable
field $\beta:M \to \mathbb{R}$, moreover $g(n,V) <0$ since $v$ is
future directed. Hence
\[
\beta\, g(n,V)=g(L_n V,V)=\frac{1}{2}L_n g(V,V) ,
\]
implies that (i), i.e. $\beta=0$, iff (iii). Since $n$ is Killing
(i) iff (ii). Let $V'$ be a representative.  By choosing a
suitable field $\alpha: M \to \mathbb{R}$, we can find another
representative $V=V'+\alpha n$ such that
\[
g(V,V)=g(V',V')+2 g(V',n) \alpha=g(V',V')-2 (\psi \dd t[v])\alpha
,
\]
is any
prescribed function, and a condition of this kind fixes the
function $\alpha$ and hence the representative. In particular $V$
can be chosen to be causal and such that Eq. (\ref{lift3}) holds
with $-g(n,V)=\psi \dd t[v]>0$.
\end{proof}

In what follows we shall be interested only in the case in which
$\mu$ is a constant throughout $E$. The reason for introducing the
function $\mu$ stands on the fact that for future directed timelike
geodesic on $M$  both $\sqrt{-g(V,V)}$ and $-g(n,V)$ are positive
constants where $V$ is the tangent vector. Then $\mu$ is independent
of the affine parametrization and it is the momentum conjugate  to
$y$ with respect to the proper time action (see section \ref{ceis}).

Note that if $\psi \dd t[v]=1$, i.e. $v(e)$ is a Newtonian frame,
then $-g(n,V)=1$.  The 0-lift corresponds to $V$ lightlike, and will
be called the {\em lightlike} or {\em light } lift of $v$. The
$\frac{1}{\mu}$-lift makes sense not only for vector fields but also
for vectors and  if $V$ is the $\frac{1}{\mu}$-lift of $v$ then
$\lambda V$, $\lambda >0$, is the $\frac{1}{\mu}$-lift of $\lambda
v$.

Let $V$ be the tangent vector to a causal curve on $M$, which
projects to a future directed curve of tangent vector $v$ (i.e.
$g(n,V)<0$). The curves on $E$ and $M$ can be parametrized with
respect to $t$, and  Eq. (\ref{lift3}) reads
\begin{equation} \label{sdc}
\dot{y}=\frac{1}{2 \psi}\,a_t(\dot{q},
\dot{q})+b_t(\dot{q})+\psi(\frac{1}{2\mu(t)^2}-\frac{1}{2}- U).
\end{equation}
As we shall see the curves for which $1/\mu$ is a constant will be
particularly interesting.

\subsection{Timelike Killing fields and conservation of energy} Given a Newtonian frame $v$ and a choice of
representative $V$ such that condition (i) holds, a gauge
transformation can be made in such a way that $L_V y=0$ or, which is
the same, $V=\frac{1}{\psi} \frac{\p}{\p t }$. Conversely, whatever
the gauge chosen the field $\frac{1}{\psi} \frac{\p}{\p t }$ is a
representative which commutes with $n=\p / \p y$. These expressions
are kept unaltered by the restricted gauge transformations for which
$\alpha$ in Eqs. (\ref{pz1})-(\ref{pz3}) does not depend on $t$.


Consider a Bargmann structure ($\psi=1$). The change of gauge
$y'=y-Ct$,  where $C$ is a constant, leaves unaltered the fields
$a_t$, $b_t$, and the expression  $n=\p/\p y$, while it changes the
function $U$ and the field $\p/\p t$
\begin{eqnarray}
U'&=&U+C ,\label{ft1}\\
\frac{\p}{\p t'}&=& \frac{\p}{\p t}+C \frac{\p}{\p y}. \label{ft2}
\end{eqnarray}
Notice that depending on the event and on the gauge chosen the
representative $\frac{\p}{\p t}$ may be timelike, lightlike or
spacelike. The previous observation is useful because it shows that
if $\frac{\p}{\p t}$ is Killing then it is not restrictive to assume
it timelike, provided the  function $U$ is bounded from below.

\begin{theorem} \label{nhu}
In a Bargmann structure let $V$, $L_n V=0$, be a representative
field for the Newtonian frame ${v}$. The field $\p/\p t$ on $M$ is
always a representative for $v$, and the section $\sigma$ can be
chosen in a way such that ${V}=\frac{\p}{\p t}$ on $M$, and
${v}=\frac{\p}{\p t}$ on $E$. With that  choice ${V}$ is Killing iff
$a_t$, $b_t$ and $U$
 are independent of $t$ (but note that if $a_t$ depends on $t$ then $V$
 is not Killing irrespective of the  choice of $\sigma$). Moreover, an (alternative) timelike
Killing representative can be found iff the potential $U$ is bounded
from below.
\end{theorem}

\begin{proof}
The first statement is trivial. The causal character of
${V}=\frac{\p}{\p t}$ is given by the sign of the function
$(2{U}+1)$ in Eisenhart's metric. Clearly if $U$ is bounded from
below a gauge transformation $y'=y-Ct$ can be found such that, due
to Eq. (\ref{ft1}), $(2U'+1)>0$, i.e. $\frac{\p}{\p t'}$ is
timelike, and, due to Eq. (\ref{ft2}), $\frac{\p}{\p t'}$ being a
linear combination of Killing fields is Killing. Every alternative
Killing representative has the form $\frac{\p}{\p t}+C(m)n$. Since
both $n$ and $\frac{\p}{\p t}$ are Killing, the Killing condition
reads, $C_{;\mu}n_{\nu}+C_{;\nu}n_{\mu}=0$ or $\dd C \otimes \dd
t+\dd t\otimes \dd C=0$ which implies that $C$ is a constant. Thus
any alternative Killing representative is obtained through the
already considered gauge transformation.
\end{proof}

Consider a Bargmann structure with a (timelike) Killing field $K$
which projects to a Newtonian frame i.e. let $L_K g=0$, $L_n K
\propto n$ and $g(K,n)=-1$. Set $L_n K =[n,K]=\beta n$. From $0=L_K
g(K,n)=g(K,L_K n)=-g(K,L_{n}K)=-\beta g(K,n)=\beta$ we conclude that
$L_{n}K=0$ and hence theorem \ref{nhu} can be applied with $V=K$.
Thus the section $\sigma$ can be chosen in such a way that
$K=\frac{\p}{\p t}$ and $a_t$, $b_t$ and $U$ do not depend on $t$.
Consider now a future directed causal curve $x(\lambda)$ on $M$,
with tangent vector  $\frac{\dd x}{\dd \lambda}$ nowhere
proportional to $n$, i.e. $g(n,\frac{\dd x}{\dd \lambda}) < 0$. It
is not difficult to check that
\begin{equation}
\frac{1}{2}\,a(\dot{q}, \dot{q})+ U=\frac{g(n,\frac{\dd x}{\dd
\lambda})g(K,\frac{\dd x}{\dd \lambda})-\frac{1}{2} g(\frac{\dd
x}{\dd \lambda},\frac{\dd x}{\dd \lambda})}{g(n,\frac{\dd x}{\dd
\lambda})^2}-\frac{1}{2}.
\end{equation}
If $x(\lambda)$ is a causal geodesic the right-hand side is a
constant of motion that does not depend on the affine
parametrization. It gives the conservation of energy on the reduced
spacetime $E$, and it is due to the additional symmetry generated by
$K$.

\section{Eisenhart's theorem} \label{ceis}

We give a proof of Eisenhart's geodesic characterization  of
classical dynamics. A spacelike version was previously obtained
using a Finslerian approach in \cite{lichnerowicz55} which
generalizes  the original theorem by Eisenhart \cite{eisenhart29}.
Our proof makes use of more standard tools and in particular makes
use of geometrically useful constructions, such as the
$\frac{1}{\mu}$-lift which allows us to find a relation between the
Eisenhart's theorem in the lightlike geodesic case and a Fermat type
theorem with lightlike target curves. It also allows us to apply the
tools of global Lorentzian geometry to the problem of the existence
of minimizers  for the classical action (see theorem \ref{cas}).

So far there have been only few applications of Eisenhart's
theorem\footnote{Benn \cite{benn06} and Szydlowski
\cite{szydlowski98} cite Eisenhart \cite{eisenhart29} while studying
the geodesics of a warped product spacetime with a metric of the
form $g_{a b}\dd q^a \dd q^b + A(q^a) \dd u^2$ which, according to
them, leads to a Newtonian dynamics on the reduced space of
coordinates $\{q^a\}$ by taking as time the affine parameter and by
letting $A^{-1}\propto \mathcal{U}$, with $\mathcal{U}$ the
Newtonian potential. However, this attribution seems incorrect. In
Eisenhart's work the crossed terms of the metric are of fundamental
importance, its determinant does not depend on the sign of
$\mathcal{U}$, and $A \propto \mathcal{U}$ \cite[p. 593 and Eq.
2.10]{eisenhart29}. Their work is more properly related to that on
Kaluza-Klein theories with a scalar field.}. Among those not treated
here there is one on the chaotic behavior of dynamical systems
\cite{pettini93}.

 We want to find out how geodesics on spacetime $M=E\times
\mathbb{R}=T\times S\times\mathbb{R} $  endowed with metric
(\ref{asd}) project on $E$. Since $n$ is a Killing vector field
$n_{\mu;\nu}+n_{\nu;\mu}=0$. By contraction with $n$ it follows,
$n^{\mu}_{\,;\nu}n^{\nu}=0$ that is, the integral lines of the
lightlike vector field are geodesics. We shall be interested on
causal geodesics not coincident with them  since these project on
single points of $E$, that is they are the events of the reduced
spacetime $E$.  If $V\ne 0$ is a future directed causal vector not
coincident with $n$ then $0>g(V,n)=-\psi\dd t[V]$. This equation
implies that the time variable $t$ increases along the projected
worldlines and can be used as a time parameter.

The causal geodesics $x(\lambda)$ on which we are interested are
the extremals of the action
\begin{eqnarray}
\int_{0}^{1}\mathcal{L}\dd \lambda' &=& \int_{0}^{1}- g(\frac{\dd
x}{\dd  \lambda'},\frac{\dd x}{\dd  \lambda'}) \dd
\lambda'=\int_{0}^{1}\{ -a_t(\frac{\dd q}{\dd \lambda'}, \frac{\dd
q}{\dd \lambda'})-2\psi b_t(\frac{\dd q}{\dd \lambda'}) \frac{\dd
t}{\dd \lambda'} \nonumber\\&& +(2 U+1)\psi^2 (\frac{\dd t}{\dd
\lambda'})^2 +2\psi \frac{\dd t}{\dd \lambda'}\frac{\dd y}{\dd
\lambda'}\} \dd \lambda' .
\end{eqnarray}
The parameter $\lambda$ is an affine parameter for the seeked
extremals. There are two conserved quantities. The momentum
conjugate to $y$
\begin{equation} \label{mu}
 \tilde{\mu}=2\psi\frac{\dd t}{\dd
\lambda}=-2 g(n,\frac{\dd x}{\dd  \lambda}) >0,
\end{equation}
where $\tilde{\mu}>0$ because the geodesic is causal, future
directed, and not coincident with a flow line of the lightlike
Killing vector field, and
\begin{eqnarray}
\mathcal{L}&=& - g(\frac{\dd x}{\dd  \lambda},\frac{\dd x}{\dd
\lambda})=-a_t(\frac{\dd q}{\dd \lambda}, \frac{\dd q}{\dd
\lambda})\nonumber\\&& \!\!\!\!\!\!\!\!\!-2\psi\, b_t(\frac{\dd
q}{\dd \lambda}) \frac{\dd t}{\dd \lambda} +(2 U+1)\psi^2 (\frac{\dd
t}{\dd \lambda})^2 +\tilde{\mu } \frac{\dd y}{\dd \lambda}=C^2 ,
\label{cc}
\end{eqnarray}
where $C>0$ for a timelike geodesic and $C=0$ for a lightlike
geodesic. The last equation can be used in place of the
Euler-Lagrange equation corresponding  to the coordinate $t$. The
Euler-Lagrange equations obtained from the variation with respect to
$q$ can be simplified by removing $\lambda$, through Eq. (\ref{mu}),
and $y$ through Eq. (\ref{cc}). The resulting equations will be
called the projected Euler-Lagrange equation as the projection
$q(t)$ on $E$ of a geodesic satisfies them. Nevertheless, we will
not need to find them explicitly. Instead, our problem will be that
of finding a direct variational formulation for them.

Note that the conserved quantities $\tilde\mu$ and $C$ depend on the
chosen affine parametrization and are fixed by the requirement that
at the endpoints $\lambda=0$ and $\lambda=1$, respectively. However,
due to the homogeneity in $\lambda$ of the Euler-Lagrange equations,
the projected Euler-Lagrange equations will be dependent only on the
already introduced ratio
\begin{equation} \label{lkm}
\frac{1}{\mu}=\frac{2C}{\tilde{\mu}}=\frac{\sqrt{- g(\frac{\dd
x}{\dd  \lambda},\frac{\dd x}{\dd  \lambda})}}{-g(n,\frac{\dd
x}{\dd  \lambda})},
\end{equation}
 and will be called $\frac{1}{\mu}\,$-Euler-Lagrange equations.
Note that $\mu \in (0,+\infty]$, where $\mu=+\infty$ iff $C=0$. One
can easily check that $\mu \ne 0$ is the momentum  conjugate to $y$
for the Lagrangian $\mathcal{L}^{1/2}$ which leads to the usual
proper time action (we did not start directly from that action
because, in this way, we include the lightlike geodesics in our
analysis).

Eq. (\ref{mu}) implies that in a Bargmann structure $t$ is an affine
parameter for the geodesic. More generally, Eq. (\ref{cc}) implies
that for any timelike geodesic
\begin{equation}
\dd \tau=\frac{\psi}{\mu}\,\dd t .
\end{equation}
where $\tau$ is the proper time.
The equation (\ref{lkm}) reads
\begin{equation} \label{lift}
y(t)=y(t_0)+ \int_{t_0}^t[\frac{1}{2 \psi}\,a_{t'}(\dot{q},
\dot{q})+b_{t'}(\dot{q})+\psi(\frac{1}{2\mu^2}-\frac{1}{2}- U)] \dd
t'.
\end{equation}

\begin{definition}
Given  $\mu \in (0,+\infty]$, a  future directed curve $(t,q(t))$ on
$E$ and a point $m_0=(t_0, q(t_0), y(t_0))$ (determined by a choice
of $y_0=y(t_0)$) on the fiber of the starting point $e_0=(t_0,
q(t_0))$, equation (\ref{lift}) defines a new future directed causal
curve on $M$, given by $(t,q(t),y(t))$, which we call the
$\frac{1}{\mu}\,$-lift of $(t,q(t))$. The 0-lift will also be called
{\em light lift}.
\end{definition}
The $\frac{1}{\mu}\,$-lift can be defined for any future directed
curve on $E$ and, by construction, it is a timelike curve on $M$ if
$0<\mu<+\infty$ and a lightlike curve on $M$ if $\mu=+\infty$. The
final endpoint $m_1= (t_1, q(t_1), y(t_1))$ of the
$\frac{1}{\mu}\,$-lift lies in the fiber of the final endpoint $e_1$
of the original curve on $E$.

The importance of the $\frac{1}{\mu}\,$-lift lies in the expression
for the coordinate $y$ of its second endpoint which clearly
resembles the classical action.

We are generalizing here the idea of light lift introduced in
 \cite{minguzzi03b} to the case of lightlike dimensional reduction.
 In \cite{minguzzi03b} in order to study the motion of a relativistic
 particle in  electromagnetic and gravitational fields, a
 Kaluza-Klein extension of the spacetime was considered (spacelike dimensional reduction). The
 additional coordinate of the light lift was proved to be related to
 the action of the particle, in complete analogy with the present case.
 The relevant difference is that here the dynamics on the base is non-relativistic
 while the bundle is a spacetime, while
 there both manifolds were  relativistic spacetimes.

We have almost proved the following
\begin{theorem}  \label{fond1}
Let $(M,g)$, $M=E \times Y=T\times S\times Y$, $Y= \mathbb{R}$, $T$
connected open subset of $\mathbb{R}$, be the  spacetime endowed
with the metric (\ref{asd}), where $a_t$, $b_t$, $U$ and $\psi$ are
(all possibly time dependent) a positive definite metric, 1-form
field,  scalar field and  positive scalar field over $S$.


Every causal geodesic, not coincident with a flow line of $n=\p/\p
y$, has a constant of motion $\frac{1}{\mu}\in [0,+\infty)$, and a
projection to $E$, $(t, q(t))$, which is a $C^2$ stationary point of
\begin{equation} \label{nja}
E_{1/\mu}[q(t)]=\int_{t_0}^{t_1}[\frac{1}{2\psi} \, a_t(\dot{q},
\dot{q})+b_t(\dot{q})-\psi(\frac{1}{2\mu^2}+\frac{1}{2}+ U)] \dd t'
,
\end{equation}
on the set $C^1_{e_0,e_1}$ of $C^1$ curves $q:[t_0,t_1] \to S$ with
fixed endpoints $q(t_0)=q_0$, $q(t_1)=q_1$.

 Moreover, let
$\frac{1}{\mu}\in  [0,+\infty)$ and let $(t, q(t))$ be a future
directed curve on $E$, then $q(t)$ is a $C^2$ stationary point of
the functional $E_{1/\mu}$ on the set $C^1_{e_0,e_1}$ iff the
$\frac{1}{\mu}\,$-lift to $M$ obtained by defining $y(t)$ as in Eq.
(\ref{lift}) is a (unparametrized) geodesic.  If $\mu\ne 0$ the
geodesic is timelike and the proper time parametrization is obtained
integrating $\dd \tau =\frac{\psi}{\mu}\dd t$. If $\frac{1}{\mu}=0$
the geodesic is lightlike and an affine parametrization is obtained
integrating $\dd \lambda=\psi \dd t$.
\end{theorem}

In order to complete the proof we have still to solve our main
problem, i.e. to prove that the $\mu$-dependent variational
principle  $\delta E_{1/\mu}[q(t)]=0$ on $E$  leads directly to the
projected $\frac{1}{\mu}$-Euler-Lagrange equations and that the
$\frac{1}{\mu}\,$-lifts of the stationary points are geodesics on
$M$.  In the next sections we shall complete the proof by
considering separately the cases $\mu=+\infty$ and $\mu <+\infty$.

Note that not every future directed causal curve on $M$ can be
regarded as a $\frac{1}{\mu}$-lift of a curve on $E$ since the
function $\frac{1}{\mu}(t)$ calculated through Eq. (\ref{lkm}) must
be a constant for this to be the case. However, any lightlike curve
on $M$ with tangent vectors nowhere proportional to $n$, can be
regarded as the light lift of its own projection  on $E$ because in
this case $\frac{1}{\mu}(t)=0$.

By calculating the Euler-Lagrange equations of the functional
$E_{1/\mu}[q(t)]$ we find that the potential $b_t$ gives rise to a
magnetic type force proportional to the curvature of the connection,
that is, exactly the Coriolis force that we expected.

A consequence of the previous theorem is the following theorem which
generalizes Eisenhart's.

\begin{theorem}   \label{fond2}
Let $\tilde{a}_t$, $b_t$ and $\mathcal{U}$ be (possibly time
dependent) positive definite metric, 1-form field and  scalar field
(potential) on the configuration space $S$. For any choice of
constant $\mu \in (0,+\infty]$,  and time dependent field $\psi>0$,
every $C^2$ stationary point $(t,q(t))$ of the classical action
($t_0<t_1$)
\begin{equation} \label{bjs0}
\int_{t_0}^{t_1} [\frac{1}{2}\tilde{a}_t(\dot{q},
\dot{q})+b_t(\dot{q})- \mathcal{U}(t,q)] \dd t' ,
\end{equation}
on the set $C^1_{e_0,e_1}$ can be regarded, up to
reparametrizations, as the projection of a causal future directed
geodesic on a $(d+1)+1$-dimensional spacetime $M=T \times S \times
Y$, $Y=\mathbb{R}$,  endowed with the metric
\begin{equation} \label{bjs1}
g_{\frac{1}{\mu},\psi}=\psi[\pi_S^{*}\tilde{a}_t-\dd t \otimes (\dd
y-\pi_S^{*} b_t)-(\dd y-\pi_S^{*} b_t) \otimes \dd t
-(2\mathcal{U}-\frac{\psi}{\mu^2}) \dd t^2 ] . \ \
\end{equation}
The geodesic is timelike if $\frac{1}{\mu}\ne 0$, in which case the
proper time along the geodesic is $\dd \tau= \frac{\psi}{\mu}\dd t$,
and lightlike if $\frac{1}{\mu}=0$, in which case an affine
parametrization is $\dd \lambda =\psi \dd t$. Moreover, the
(unparametrized) geodesic is the $\frac{1}{\mu}$-lift of the
stationary point on $E$, i.e. it is obtained by setting
\begin{equation} \label{bjs2}
y(t)=y(t_0)+ \int_{t_0}^t[\frac{1}{2}\tilde{a}_t(\dot{q},
\dot{q})+b_t(\dot{q})+(\frac{\psi}{\mu^2}- \mathcal{U})] \dd t'.
\end{equation}
Conversely, every  future directed geodesic on
$(M,g_{\frac{1}{\mu},\psi})$ which is, up to reparametrizations, the
$\frac{1}{\mu}\,$-lift of its own projection $(t,q(t))$ on $E$ is
such that $q(t)$ is a $C^2$ stationary point of the classical action
on the set $C^1_{e_0,e_1}$. In particular, every lightlike geodesic
on $(M,g_{0,\psi})$, not coincident with a flow line of $n=\p/\p y$,
projects on a stationary point of the classical action and every
stationary point can be obtained in this way.
\end{theorem}

Recall that the stationary points of the action functional
(\ref{bjs0}) do not depend on the choice of zero for the potential.
Thus in Eqs. (\ref{bjs1})-(\ref{bjs2}) $\mathcal{U}$ can be replaced
with $\mathcal{U}+C$ where $C\in \mathbb{R}$ is an arbitrary
constant, however note that the metric so obtained is actually
isometric to the previous one since the constant can be removed with
a change of coordinate $y \to y-Ct$.

Theorem \ref{fond2} follows immediately from theorem \ref{fond1}
once the identification
\[
\mathcal{U}=\psi(\frac{1}{\mu^2}+\frac{1}{2}+U)+const.
\]
has been made. In particular for a Bargmann structure, $\psi=1$ and
$U$ is then a potential equivalent to $\mathcal{U}$.

\section{The proof} \label{sfg}

The proof of theorem \ref{fond1}, especially in the lightlike case,
is particularly interesting and will be needed for a deeped
understanding of the next section.

\subsection{The lightlike geodesic case, $\mu=+\infty$} \label{cd2}
\label{ceis2} We have to prove that the projection  of a lightlike
geodesic $x(\lambda)$ on $M$, not coincident with a flow line of
$n$, is, once parametrized with respect to $t$, a stationary point
of the functional
\[
E_{0}[q(t)]=\int_{t_0}^{t_1}[\frac{1}{2\psi}a_t(\dot{q},
\dot{q})+b_t(\dot{q})- (U+\frac{1}{2})\psi ]\dd t' ,
\]
and conversely that the light lift of a stationary point of this
functional is, with a suitable parametrization, a lightlike geodesic
on $M$. In order to reach this result we recall that the
unparametrized lightlike geodesics do not change under conformal
changes of the metric \cite[Appendix D]{wald84}. As a consequence
the lightlike geodesic $x(\lambda)$ on $M$ can be suitably
reparametrized to become a stationary point of the action
\begin{equation} \label{xsa}
\!\!\!\!\!\!\!\!\!\!\!\!\!\!\!\!\!\!\!\!\!\!\int_{\tilde\lambda_0}^{\tilde\lambda_1}
\!\frac{1}{2\psi} \,g(\frac{\dd x}{\dd  \lambda'},\frac{\dd x}{\dd
\lambda'})\, \dd \tilde\lambda'
=\int_{\tilde\lambda_0}^{\tilde\lambda_1}\!\! [\frac{1}{2\psi}\,
a_t(\dot{q}, \dot{q})+b_t(\dot{q}) - (U+\frac{1}{2})\psi-\dot{y}]
(\frac{\dd t}{\dd \tilde\lambda'})^2 \dd \tilde\lambda' .
\end{equation}

%
Conversely, if a causal curve $x(\tilde\lambda)$ with tangent
vectors nowhere proportional to $n$  satisfies the previous
variational principle then it can be reparametrized to become a
lightlike geodesic for $(M,g)$. Let $x(\tilde\lambda,r)$, $r\ge0$,
be a $(C^1)$ variation with fixed endpoints  of a causal curve
$x(\tilde\lambda)=x(\tilde\lambda,0)$, with tangent vectors nowhere
proportional to $n$. The continuous function in $(\tilde\lambda,r)$,
$g(n,x_{\tilde\lambda}(\tilde\lambda,r))$, where
$x_{\tilde\lambda}(\tilde\lambda,r)$ is the longitudinal tangent
vector, is negative for $r=0$, i.e. over $x(\tilde\lambda)$. Since
the curve $x(\tilde\lambda)$ is defined over a compact there is an
$\epsilon>0$ such that for $r<\epsilon$ the longitudinal curves of
the variation have tangent vectors nowhere proportional to $n$.
Therefore, it is not restrictive to assume that the longitudinal
curves do not have tangent vectors proportional to $n$. As a
consequence the curves of the variation can be reparametrized with
respect to $t$ because $\dd t/\dd \tilde\lambda
=-\psi^{-1}g(n,x_{\tilde\lambda})>0$. Thus there are functions
$\hat{x}(t,r)$ and $t(\lambda,r)$ such that
$x(\lambda,r)=\hat{x}(t(\lambda,r),r)$. Conversely, for any pair of
functions $\hat{x}(t,r)$ and $t(\lambda,r)$ the previous equation
gives a  variation $x(\lambda,r)$, the only conditions to be imposed
on $\hat{x}(t,r)$ and $t(\lambda,r)$ are as follows,
$x(\lambda)=\hat{x}(t(\lambda,0),0)$ where $\hat{x}(t,0)$ is the
reparametrization with $t$ of $x(\lambda)$ and $t(\lambda,0)$ gives
the dependence between the two parametrization on the same curve,
and finally, $t(\lambda_0,r)=t_0$, $t(\lambda_1,r)=t_1$,
$\hat{x}(t_0,r)=x(\lambda_0)$, $\hat{x}(t_1,r)=x(\lambda_1)$.

 The particular form of $\hat{x}(t,r)$ and
$t(\lambda,r)$ selects the variational field
 which reads $x_{r}\vert_{r=0}=\hat{x}_{t}
 t_{r}\vert_{r=0}+\hat{x}_{r}\vert_{r=0}$. The idea is first to fix
 $t(\lambda,r)=t(\lambda,0)$ and find the Euler-Lagrange equations
 corresponding to the freedom induced by variations of the form
 $x(\lambda,r)=\hat{x}(t(\lambda,0),r)$ and then to fix
 $\hat{x}(t,r)=\hat{x}(t,0)$ and find the Euler-Lagrange equation
 corresponding to the freedom induced by variations of the form
 $x(\lambda,r)=\hat{x}(t(\lambda,r),0)$.

Note that $\hat{x}(t,r)=(t,q(t,r),y(t,r))$. The variation
$\hat{x}(t,r)$ which keeps $q(t,r)=q(t)$ but changes $y(t)$ gives
the Euler-Lagrange equation
\begin{equation} \label{cgi}
\dd t/\dd \tilde\lambda =cnst.>0,
\end{equation} that is, if $x(\lambda)$ is a lightlike
geodesic for $(M,g)$ then $t$ is an affine parameter for the same
curve regarded as a lightlike geodesic of $(M,g/\psi)$. The relation
between $\lambda$ and $t$ is then $\dd \lambda= K \psi\dd t$ for a
suitable constant $K \in \mathbb{R}^{+}$ (see \cite[Eq.
(D6)]{wald84}).

 Using Eq. (\ref{cgi}) it is easily shown that the variations which
keep $y(t,r)=y(t)$ but change $q(t)$ lead to the same Euler-Lagrange
equations of functional $E_0$. Indeed, the term $-\dot{y}(\frac{\dd
t}{\dd \tilde\lambda'})^2$ does not change under the said variations
and does not contribute to the Euler-Lagrange equations. Also the
factor $(\frac{\dd t}{\dd \tilde\lambda'})^2$ in the remaining term
$[\frac{1}{2\psi}\, a_t(\dot{q}, \dot{q})+b_t(\dot{q}) -
(U+\frac{1}{2})\psi] (\frac{\dd t}{\dd \tilde\lambda'})^2$ does not
play any role because under integration by parts it gets
differentiated with respect to $t$ and vanishes due to Eq.
(\ref{cgi}). As a result the Euler-Lagrange equations corresponding
to the said variations are those of $E_0$.

Finally, the variation $x(\lambda,r)=\hat{x}(t(\lambda,r),0)$  gives
Eq. (\ref{lift}), hence every lightlike geodesic not coincident with
a flow line of $n$ is the light lift of a stationary point of $E_0$
parametrized so that $\dd \lambda= K \psi\dd t$ for a suitable
constant $K \in \mathbb{R}^{+}$. Conversely, the light lift of a
stationary point of $E_0$ satisfies all the Euler-Lagrange equations
of functional (\ref{xsa}), i.e. it is a lightlike geodesic in
$(M,g/\psi)$ and once parametrized with $\dd \lambda=\psi \dd t$
becomes a lightlike geodesic for $(M,g)$.

%

\subsection{The timelike geodesic case, $\mu<+\infty$} \label{ceis4}

In this case we are interested in the projection of timelike
geodesics on $M$. It is convenient to consider the variational
principle
\begin{equation}
\delta\int_{0}^{1}\mathcal{L}^{1/2} \dd \lambda=0 ,
\end{equation}
over the set of $C^2$ timelike curves with fixed endpoints. The
action is the usual proper time.
 The cyclic variable $y$ can be removed from the
variational principle using Routh's reduction \cite{marsden99} (see
also \cite{lichnerowicz55} for the application to Lagrangians
homogeneous of first degree in the velocities). The Routhian is a
reduced Lagrangian obtained from the original Lagrangian by making a
Legendre transform with respect to the velocities of the cyclic
variables that have to be  removed, and by considering the conserved
conjugate momenta as constants. In our case we want to remove the
variable $y$ and its velocity $\dd{y}/\dd \lambda$ so as to obtain a
variational principle in the quotient space $E$. The conjugate
momentum is
\begin{eqnarray*}
\mu&=&\mathcal{L}^{-1/2}\psi \frac{\dd t}{\dd \lambda}\\
&=&\frac{\psi}{\{-a_t(\dot{q},\dot{q})-2\psi b_t(
\dot{q})+(2U+1)\psi^2+2\psi \dot{y} \}^{1/2}},
\end{eqnarray*}
from which we obtain that $y(t)$ is given by Eq. (\ref{lift}). The
Routhian is
\begin{eqnarray*}
R&=&\mu \frac{\dd y}{\dd \lambda}-\mathcal{L}^{1/2}=[\mu
\dot{y}-\frac{\psi}{\mu} ]\frac{\dd t}{\dd
\lambda} \\
&=&[\frac{1}{2\psi}a_t(\dot{q}, \dot{q})+
b_t(\dot{q})-\psi(\frac{1}{2\mu^2}+ \frac{1}{2}+ U)]\mu\frac{\dd
t}{\dd \lambda}  ,
\end{eqnarray*}
and, up to a constant factor, the reduced action becomes
\[
\frac{1}{\mu} \int_{0}^{1} R \dd
\lambda=\int_{t_0}^{t_1}[\frac{1}{2\psi}a_t(\dot{q},
\dot{q})+b_t(\dot{q})-\psi(\frac{1}{2\mu^2}+\frac{1}{2}+ U)] \dd t'.
\]
The proof of theorem \ref{fond1} and hence of theorem \ref{fond2}
are complete.

\section{The geometric interpretation of the lightlike geodesic case: Fermat's principle and
Bolza's problem} \label{ceis3}

In this section $(M,g)$ is the Eisenhart spacetime $M=T\times
S\times \mathbb{R}$, $T$ open connected subset of $\mathbb{R}$, with
$g$ given by Eq. (\ref{asd}) with (all possibly time dependent)
$a_t$, a positive definite metric, $b_t$, a 1-form field, $U$, a
scalar field and $\psi>0$, a positive scalar field.

 Let us
consider points $e_0, e_1 \in E=T\times S$ with $e_1$ in the future
of $e_0$, $t_1>t_0$. Let $m_0=(e_0,y_0)$ be an event on the fiber of
$e_0$ in the bundle $\pi: M \to E$. Consider the set
$\mathcal{N}_{m_0,e_1}$ of the (unuparametrized) $C^1$ lightlike
curves connecting $m_0$ to $e_1$'s fiber. The coordinate $y_1$ of
the final endpoint $m_1=(e_1,y_1)$ can be regarded as a functional
on $\mathcal{N}_{m_0,e_1}$. Note that the parametrization $\lambda$
of the curves considered will play no role in the discussion and can
always be taken such that $\lambda \in [0,1]$. Note also that the
definition of stationary point for the functional $y_1$ is not
affected by changes of the section $\sigma$, as this implies only
the addition of a constant.

\begin{lemma} \label{vfg} If $g(n,\p_{\lambda}) = 0$ at a point of a curve $x \in \mathcal{N}_{m_0,e_1}$
then a
 lightlike variation $x(\lambda,r)$, $r \ge 0$, with longitudinal curves belonging to $\mathcal{N}_{m_0,e_1}$, can be
constructed such that $\dd y_1/\dd r>0$. In particular, a stationary
point $x \in \mathcal{N}_{m_0,e_1}$, of the arrival coordinate $y_1$
on $\mathcal{N}_{m_0,e_1}$ satisfies $g(n,\p_{\lambda}) \ne 0$ at
every point. The same inequality holds for any lightlike geodesic in
$\mathcal{N}_{m_0,e_1}$.
\end{lemma}

\begin{proof}
That the statement holds for any geodesic follows from the fact that
$g(n,\p_{\lambda})(\bar{\lambda})=0$, $\bar{\lambda} \in [0,1]$,
would imply that the lightlike geodesic $x(\lambda)$ is tangent to
$n$ at a point and hence coincides with the lightlike geodesic
generated by $n$, which is impossible as the initial and final
endpoints of $x(\lambda)$ project on distinct points of $E$.

Now, consider a curve $x(\lambda) \in \mathcal{N}_{m_0,e_1}$. The
idea is that if $g(n,\p_{\lambda})(\bar{\lambda})=0$, $\bar{\lambda}
\in (0,1]$, then there is a lightlike variation of $x(\lambda)$,
$\lambda \in[0,\bar{\lambda}]$, which keeps the projection of the
endpoint $\bar{e}=e(\bar\lambda)$ fixed but changes the value of
$\bar{y}=y(\bar\lambda)$ linearly in the variational parameter. Then
the variation of the whole curve $x(\lambda)$ is obtained by gluing
the said variation up to $\bar\lambda$ to the Lie translation of
$x(\lambda)$ along the fibers for $\lambda>\bar{\lambda}$.

A similar idea is followed if $\bar\lambda=0$. In this case it
suffices to follow the geodesic generated by $n$ at $m_0$ and then
to move on the Lie translation of $x(\lambda)$ along the fiber.

 If
$g(n,\p_{\lambda})(\bar{\lambda})=0$, $\bar{\lambda} \in (0,1]$, we
want to show that there is a lightlike variation of $x(\lambda)$,
$\lambda \in[0,\bar{\lambda}]$, which keeps the projection of the
endpoint $\bar{e}=e(\bar\lambda)$ fixed but changes the value of
$\bar{y}=y(\bar\lambda)$. To this end, let $\eta(y')$ be the
integral line of the geodesic generated by $n$ and such that
$\eta(0)=x(\bar\lambda)$, i.e. $y'=y-\bar{y}$, and let $k$ be the
constant such that $\frac{\dd x}{\dd \lambda}(\bar{\lambda})=k n$.
Consider the following variation $x(\lambda, r)$ for $r\ge 0$
\begin{eqnarray*}
x(\lambda,r)&=&x((1+r) \lambda), \qquad 0\le \lambda \le
\frac{\bar\lambda}{1+r} \\
x(\lambda,r)&=& \eta(k((1+r)\lambda-\bar\lambda)) , \qquad
\frac{\bar\lambda}{1+r} \le \lambda \le \bar\lambda
\end{eqnarray*}
it consists in extending the lightlike curve along the geodesic and
in rescaling the parameter. Clearly the final endpoint being
$\eta(kr \bar{\lambda})$ has a coordinate $y$ of value $\bar{y}+k
\bar{\lambda} r$, which increases linearly with $r$.

\end{proof}
%

\begin{lemma} \label{bhe}
The lightlike geodesics connecting $m_0=(e_0,t_0)$ to $e_1$'s fiber,
$t_0<t_1$, are stationary points for the arrival coordinate
functional $y_1$ on the set $\mathcal{N}_{m_0,e_1}$.
\end{lemma}

\begin{proof}
Let $x(\lambda,r)$,  be a variation of $x(\lambda,0)=x(\lambda)$
made of curves belonging to $\mathcal{N}_{m_0,e_1}$. The variational
field $\p_r$ vanishes at $\lambda=0$ while
$\p_{r}\vert_{(\lambda,r)=(1,0)}=\frac{\dd y}{\dd r} n$, hence the
functional given by the function $y$ on the final endpoint has a
stationary point at $x(\lambda,0)$ iff the variational field
vanishes at the final endpoint. Since
$g(\p_{\lambda},\p_{\lambda})=0$,
\[
0=\p_{r} g(\p_{\lambda},\p_{\lambda})=2 \p_{\lambda}
g(\p_{\lambda},\p_{r})-2g(\nabla_{\p_{\lambda}} \p_{\lambda},
\p_{r}) ,
\]
If $x(\lambda)$ is a geodesic, up to reparametrizations,
$\nabla_{\p_{\lambda}} \p_{\lambda}=f \p_{\lambda}$ for a certain
function $f(\lambda)$. Thus
$g(\p_{\lambda},\p_{r})\vert_{r=0}=C\exp\{\int^{\lambda}_0 f \dd
\lambda' \}$, and since $\p_r\vert_{\lambda=0}=0$, we have $C=0$.
Finally,
\[
0=g(\p_{\lambda},\p_{r})\vert_{(\lambda,r)=(1,0)}=\frac{\dd y}{\dd
r} \, g(\p_{\lambda},n)\vert_{(\lambda,r)=(1,0)}.
\]
But $g(\p_{\lambda},n)\vert_{(\lambda,s)=(1,0)}\ne 0$ by lemma
\ref{vfg}, hence if $x(\lambda,0)$ is a lightlike geodesic then it
is a stationary point for the functional $y_1$.
\end{proof}

Consider the set $\tilde{\mathcal{N}}_{m_0,e_1} \subset
\mathcal{N}_{m_0,e_1}$, made of those curves with tangent vectors
newhere proportional to $n$. By  lemma \ref{vfg} the stationary
points for functional $y_1$, and the geodesics on
$\mathcal{N}_{m_0,e_1}$ belong in fact to
$\tilde{\mathcal{N}}_{m_0,e_1}$. Note also that if $x(\lambda)$ is a
stationary point for variations having longitudinal curves in
$\mathcal{N}_{m_0,e_1}$ then the same is true for variations
restricted to $\tilde{\mathcal{N}}_{m_0,e_1}$.

Let $(t,q(t))$ be a ($C^1$) future directed curve connecting $e_0$
to $e_1$  and let $x(\lambda)$, $\lambda \in [0,1]$, be its
 light lift of starting event $m_0=(e_0,y_0)$ and final endpoint $m_1=(e_1,y_1)$ on $e_1$'s
fiber. Here the parametrization $\lambda$ is not important as long
as $\dd t/\dd \lambda>0$.

The light lift belongs to $\tilde{\mathcal{N}}_{m_0,e_1}$ and every
curve in $\tilde{\mathcal{N}}_{m_0,e_1}$ is the light lift of a
future directed connecting curve on $E$, where $y_1$ is given by Eq.
(\ref{lift}) with $\frac{1}{\mu}=0$. As a consequence we have proved
again the `if' part of

\begin{theorem} \label{cxz}Let ${a}_t$, $b_t$, $U$ and $\psi>0$ be (possibly time dependent) positive definite metric,
 1-form field and scalar fields
 on the $d$-dimensional  configuration space $S$. The (classical) action functional
\begin{equation} \label{lift2}
\{y(t_1)-y(t_0)\}[q(t)]=
\int_{t_0}^{t_1}[\frac{1}{2\psi}a_t(\dot{q},
\dot{q})+b_t(\dot{q})-(U+\frac{1}{2})\psi] \dd t' ,
\end{equation}
is stationary on $q(t)$ if and only if the light lift
$(t,q(t),y(t))$ on $M$ endowed with the metric (\ref{asd}) is a
(unparametrized) lightlike geodesic. In particular the
$0$-Euler-Lagrange equations are the Euler-Lagrange equations of
this functional.
\end{theorem}

The `only if' part has been proved through the argument of
subsection \ref{cd2}, indeed we can use the fact that for
$\frac{1}{\mu}=0$ the expressions for functional $E_0$ and $y_1$
differ only by a constant (compare Eq. (\ref{nja}) with Eq.
(\ref{lift})). Incidentally, by proving the `only if' part we showed
that the stationary point of the functional $y_1$ on the set
$\tilde{\mathcal{N}}_{m_0,e_1}$ is a geodesic and hence, by lemma
\ref{bhe}, that it is also a stationary point with respect to
variations in the larger space of longitudinal curves
$\mathcal{N}_{m_0,e_1}$. Summarizing we proved that, given the
functional $y_1$, any stationary point on the set
$\mathcal{N}_{m_0,e_1}$ belongs to $\tilde{\mathcal{N}}_{m_0,e_1}$,
and it is a stationary point for the restricted variations belonging
to $\tilde{\mathcal{N}}_{m_0,e_1}$ (lemma \ref{vfg}). These last
stationary points are indeed geodesics (theorem \ref{cxz}), and
hence belong to $\mathcal{N}_{m_0,e_1}$ and are stationary points
with respect to the enlarged set of variations (lemma \ref{bhe}).
The conclusion is that

\begin{theorem} \label{ferm}
The stationary points of the functional $y_1$ over the set
$\mathcal{N}_{m_0,e_1}$ of lightlike $C^1$ curves connecting
$m_0=(e_0,y_0)$ to $e_1$'s fiber, $t_0<t_1$, are the lightlike
geodesics belonging to $\mathcal{N}_{m_0,e_1}$.
\end{theorem}

This is a modified Fermat type theorem in which the target curve is
not timelike but lightlike. It is likely that the theorem could be
generalized to arbitrary spacetimes, not necessarily with a
lightlike Killing vector field, as it has been proved in the
timelike case \cite{kovner90,perlick90}.


\begin{figure}
\centering \psfrag{P}{$\pi$} \psfrag{XD}{$y$} \psfrag{T}{$t$}
\psfrag{M}{$M$} \psfrag{N}{$n$} \psfrag{S}{$S$} \psfrag{Q}{$E$}
\psfrag{NT}{$N_{t}$} \psfrag{ST}{$S_t$} \psfrag{Ga}{$e(t)$}
\psfrag{Gam}{$x(t)$} \psfrag{Q0}{$e_0$} \psfrag{Q1}{$e_1$}
\psfrag{M0}{$m_0$} \psfrag{J}{$J^{+}(m_0)\cap \pi^{-1}(e_1)$}
\includegraphics[width=7.5cm]{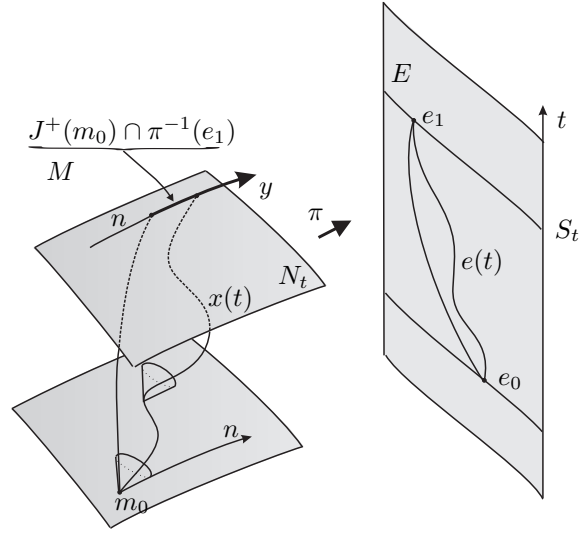}
\caption{Every future directed curve $e(t)=(t,q(t))$ connecting two
events $e_0$ and $e_1$ in $E$ is the projection of its light lift
$x(t)$ i.e. a lightlike, and hence causal, curve on $M$ which starts
from a given event $m_0$ on $e_0$'s fiber and ends on $e_1$'s fiber.
The coordinate $y$ on the light lift second endpoint equals, up to
an additive constant, the classical action functional (\ref{cxs}).
By a generalized Fermat's theorem the lightlike geodesics  are the
stationary points of this modified `time'-of-arrival functional, for
 variations restricted to lightlike curves (see theorem \ref{ferm}). Moreover, if
the action (\ref{cxs}) is bounded from below and the spacetime
$(M,g)$ is causally simple then there is a lightlike geodesic that
reaches a minimum for the coordinate $y$, and hence its projection
is an absolute minimum of the classical action (see theorem
\ref{cas}).} \label{fermat}
\end{figure}

\subsection{The Bolza problem} The Lagrange problem, sometimes referred to as the Bolza
problem, \cite{bliss46,ioffe79,cesari83,ekeland96,bolle99}  asks to
determine under which conditions the action functional has a
stationary point and in particular whether it has a minimum. A
classical result is Tonelli's theorem \cite{cesari83,mane93} which
states that a minimum exists provided $a_t$ is independent of time,
$b_t=0$, $(S,a)$ is a complete manifold, and the Lagrangian $L$ is
superlinear, that is, for all $A\in \mathbb{R}$ there is a $B \in
\mathbb{R}$ such that $L(q,\dot{q},t)\ge A
\sqrt{a(\dot{q},\dot{q})}-B$ for all $t$, $q$, $\dot{q}$.
Unfortunately, these requirements are too restrictive, they imply
for instance that the potential is bounded from above. For this
reason other theorems have been given in the literature which
replace superlinearity with other suitable growth conditions
\cite{ekeland96,gliklikh97,bolle99,candela03b} on $U$ or $\nabla U$.

A more general approach is followed here where these growth
conditions are regarded as accessory conditions needed to prove the
causal simplicity \cite{minguzzi06c} of the spacetime $(M,g)$ under
which the existence of a minimum for the classical action follows.
Indeed, remarkably, we are able to prove the following theorem which
reduces the Bolza problem for the classical action to an often
simpler problem in global Lorentzian geometry. It also summarizes
some relevant results obtained in the lightlike geodesic case
$\frac{1}{\mu}=0$.


\begin{theorem} \label{cas}
Let $S$ be a $d$-dimensional manifold endowed with the (all possibly
time dependent) positive definite metric $a_t$,  1-form field $b_t$
and potential function $\mathcal{U}$ (all $C^r$, $r \ge 1$). On the
classical spacetime $E=T \times S $ let $t$ be the time coordinate
and let $e_0=(t_0,q_0)$ and $e_1=(t_1,q_1)$ be events, the latter in
the future of the former i.e. $t_1>t_0$. Consider the classical
action functional
\begin{equation} \label{cxs}
I_{e_0,e_1}[q]=\int_{t_0}^{t_1} [\frac{1}{2} a_t(\dot{q},\dot{q})
+b_t(\dot{q})-\mathcal{U}(t,q)] \dd t,
\end{equation}
on the space $C^1_{e_0,e_1}$ of $C^1$ curves $q:[t_0,t_1] \to S$
with fixed endpoints $q(t_0)=q_0$, $q(t_1)=q_1$. Construct the
spacetime $M=E\times Y=T\times S\times Y$, $Y=\mathbb{R}$, and
denote with
 $\pi_S:M \to S$,
  the canonical projection. Assign to $M$ the Lorentzian
metric
\begin{equation} \label{me2}
\!\!\!\!\!\!\!\!\!g=\pi_{S}^{*}a_t-\dd t\otimes (\dd y-\pi_{S}^{*}
b_t) -(\dd y-\pi_{S}^{*} b_t) \otimes \dd t - 2 \mathcal{U} \, \dd
t^2.
\end{equation}
\begin{itemize}
\item[(a)] Every lightlike geodesic of $(M,g)$ not coincident with a flow
line of $n=\p/\p y$, admits as affine parameter the function $t$,
and for any such curve $x(t)=(t,q(t),y(t))$, the function $q(t)$ is
a $C^2$ stationary point of functional (\ref{cxs}) on
$C^1_{e_0,e_1}$ and $x(t)$ is the light lift of $e(t)=(t,q(t))$,
that is
\[y(t)=y(0)+ I_{e_0,e(t)}[q\vert_{[0,t]}].\]
Conversely, given a $C^2$ stationary point $q(t)$ of functional
(\ref{cxs}) on $C^1_{e_0,e_1}$, the light lift $x(t)=(t,q(t),y(0)+
I_{e_0,e(t)}[q\vert_{[0,t]}])$ is an affinely parametrized lightlike
geodesic of $(M,g)$ necessarily not coincident with a flow line of
$n$.

\item[(b)] The spacetime $(M,g)$ is causally simple if and only if the
following two properties hold
\begin{itemize}
\item[(i)] The function $\mathcal{I}: E \times E \to
[-\infty,+\infty]$ given by
\begin{eqnarray*}
\!\!\!\!\!\!\!\!\!\!\!\!\!\!\!\!\!\!\!\!\!\!\!\!\mathcal{I}(e_0,e_1)&=\inf_{q
\in C^1_{e_0,e_1}}\!\!\!
I_{e_0,e_1}[q], \qquad &\textrm{for} \ t_0<t_1, \\
\!\!\!\!\!\!\!\!\!\!\!\!\!\!\!\!\!\!\!\!\!\!\!\!\mathcal{I}(e_0,e_1)&=+\infty,
\qquad &\textrm{for} \ (t_0=t_1
\textrm{ and } q_0\ne q_1) \textrm{ or } t_0>t_1, \\
\!\!\!\!\!\!\!\!\!\!\!\!\!\!\!\!\!\!\!\!\!\!\!\!\mathcal{I}(e_0,e_1)&=
0, \qquad &\textrm{for} \ t_0=t_1 \textrm{ and } q_0= q_1,
\end{eqnarray*}
is lower semi-continuous.
\item[(ii)] If $\mathcal{I}(e_0,e_1)$, $t_0<t_1$, is finite then the functional $I_{e_0,e_1}[q]$ attains its
minimum  at a certain (not necessarily unique) $\bar{q}(t) \in
C^1_{e_0,e_1}$, i.e. $I_{e_0,e_1}[\bar{q}]=\mathcal{I}(e_0,e_1)$.
\end{itemize}
\item[(c)] Whether or not $(M,g)$ is causally simple, each
minimizer $\bar{q}(t)\in C^1_{e_0,e_1}$, is not only $C^1$ but also
$C^{r+1}$ differentiable and hence satisfies the Euler-Lagrange
equations of $I_{e_0,e_1}$.
\end{itemize}
\end{theorem}

\begin{proof}
The statement (a) is nothing but theorem \ref{fond2} for
$\frac{1}{\mu}=0$ and $\psi=1$. Thus we have only to prove (b).

 Let $\mathcal{C}_{m_0,e_1}$ be the space of $C^1$
causal curves connecting $m_0$ to $e_1$'s fiber, with $t_0 <t_1$
(see figure \ref{fermat}). Because, of the translational invariance
$m_0$ can be arbitrarily chosen. Consider a causal curve $x(\lambda)
\in \mathcal{C}_{m_0,e_1}$, $\lambda \in [0,1]$, which is timelike
at some point. By \cite[Prop. 4.5.10]{hawking73} there is a timelike
curve $\tilde{x}(\lambda)$ with the same endpoints $m_0$ and
$m_1=(e_1,y_1[x(\lambda)])$. It is easy to prove using Eq.
(\ref{sdc}) that, since $\frac{1}{\mu}>0$ all over
$\tilde{x}(\lambda)$, the coordinate $y$ evaluated at its final
endpoint (which coincides with the final endpoint of $x(\lambda)$)
is higher than the same coordinate evaluated at the final endpoint
of the light lift of the projection $\tilde{e}(\lambda)$ of
$\tilde{x}(\lambda)$.

Thus for every connecting curve which stays in
$\mathcal{C}_{m_0,e_1}$, there is another in
$\mathcal{N}_{m_0,e_1}\subset \mathcal{C}_{m_0,e_1}$ with an equal
or lower value of the functional $y_1$. Thus
\[
\inf_{x \in {\mathcal{N}}_{m_0,e_1}} y_{1}[x]=\inf_{x \in
{\mathcal{C}}_{m_0,e_1}} y_{1}[x] .
\]
Consider now a curve $x \in {\mathcal{N}}_{m_0,e_1}$, such that
$g(n, \dd x/\dd \lambda)=0$ at some point. It can not be  $g(n, \dd
x/\dd \lambda)=0$ everywhere otherwise $x$ would coincide with a
segment of geodesic generated  by $n$, and its projection would not
connect two different points $e_0$ and $e_1$. Thus it is not a
lightlike geodesic and by \cite[Prop. 4.5.10]{hawking73} there is a
timelike curve $\hat{x}$ with the same endpoints $m_0$ and
$m_1=(e_1,y_1[x(\lambda)])$, and finally the light lift of the
projection $\hat{e}$ gives a lightlike curve with a lower value of
the coordinate $y_1$ of the second enpoint. Thus for every
connecting curve which stays in $\mathcal{N}_{m_0,e_1}$, there is
another in $\tilde{\mathcal{N}}_{m_0,e_1}\subset
\mathcal{N}_{m_0,e_1}$ with an equal or lower value of the
functional $y_1$. Thus
\[
\inf_{x \in {\mathcal{N}}_{m_0,e_1}} y_{1}[x]=\inf_{x \in
\tilde{\mathcal{N}}_{m_0,e_1}} y_{1}[x] .
\]
But every curve in $\tilde{\mathcal{N}}_{m_0,e_1}$ is the light lift
of its projection i.e. of some curve $e(t)=(t,q(t))$, $q \in
C^{1}_{e_0,e_1}$. As a consequence,
\[
\inf_{x \in \tilde{\mathcal{N}}_{m_0,e_1}} y_{1}[x]=y_0+\inf_{q\in
C^{1}_{e_0,e_1} }\!\!\!\!I_{e_0,e_1}[q],
\]
and finally,
\begin{equation} \label{kjh}
\inf_{x \in {\mathcal{C}}_{m_0,e_1}} y_{1}[x]=y_0+\inf_{q\in
C^{1}_{e_0,e_1} } \!\!\!\!I_{e_0,e_1}[q]=y_0+\mathcal{I}(e_0,e_1) .
\end{equation}

We are ready to prove (c). Since $I^{+}$ is open every $C^1$
absolute minimum $\bar{q}$, must have a light lift which connects
$m_0$ to a point in $\dot{J}^{+}(m_0)=E^{+}(m_0)$, and hence the
light lift is a maximal lightlike geodesic. The geodesic equation
depends on the first derivatives of the given fields $b_t$, $a_t$,
$\mathcal{U}$, and on the second derivative of the coordinates.
Hence the geodesic has second derivatives which are $C^{r-1}$
differentiable, and the projection $\bar{q}(t)$ of the geodesic is
$C^{r+1}$ differentiable.

Let us assume $(M,g)$ is causally simple and prove property (ii). If
the action $I_{e_0,e_1}[q(t)]$ is bounded from below then there is
an event on $e_1$'s fiber,
\[ \bar{m}_1=(e_1,\bar{y}_1)=\Big( e_1,\,y_0+
\inf_{q\in C^{1}_{e_0,e_1} }\!\!\!\! I_{e_0,e_1}[q]\Big) ,
\] which belongs to $\dot{J}^{+}(m_0)$, the boundary of the causal
future of $m_0$. If $(M,g)$ is causally simple, as it is well known
\cite{hawking73}, $\dot{J}^{+}(m_0)=E^{+}(m_0)$, that is, there is a
(maximal but not necessarily unique) lightlike geodesic $\bar{x}$
connecting $m_0$ and $\bar{m}_1$. Since it is lightlike, it is the
light lift of its own projection $(t,\bar{q}(t))$, and therefore,
$\bar{q}(t)$ is an absolute minimum for the classical action.

Let us assume that $(M,g)$ is causally simple and prove property
(i). We can assume that $\mathcal{I}(e_{0},e_{1})\ne-\infty$
otherwise the statement is trivial at $(e_0,e_1)$.

Let $t_0<t_1$ and assume by contradiction that $\mathcal{I}$ is not
lower semi-continuous at $(e_{0},e_{1})\in E \times E$,  then there
is an $\epsilon>0$ and a sequence $(e_{0k},e_{1k}) \to (e_0,e_1)$
such that  $ \mathcal{I}(e_{0k},e_{1k})<
\mathcal{I}(e_{0},e_{1})-\epsilon$ (note that for $t_0<t_1$,
$\mathcal{I}(e_{0},e_{1})$ is bounded from above). Without loss of
generality we can assume $t_{0k}<t_{1k}$.

Fixed $y_0 \in \mathbb{R}$, set $m_{0k}=(e_{0k},y_0)$,
$m_0=(e_0,y_0)$, so that $m_{0k} \to m_0$. We have shown that for
any $q \in C^1_{e_{0k},e_{1k}}$, the event
\[
(e_{1k},y_0+I_{e_{0k},e_{1k}}[q])
\]
belongs to $J^{+}(m_{0k})$ and hence the same holds for
\[
(e_{1k}, y_0+I_{e_{0k},e_{1k}}[q]+C)
\]
where $C$ is an arbitrary positive constant (simply reach the
previous point and move along the geodesic generated by $n$ in the
forward direction). For $k$ sufficiently large  there is a $q \in
C^1_{e_{0k},e_{1k}}$ such that $I_{e_{0k},e_{1k}}[q]<
\mathcal{I}(e_{0},e_{1})-\epsilon$.  Finally, defined $y_{1k}$ and
$y_1$, as $y_{1k}=y_1=y_0+\mathcal{I}(e_{0},e_{1})
-\frac{\epsilon}{2}$, we obtain that
\[
m_{1k}=(e_{1k},y_{1k})=(e_{1k}, y_0+\mathcal{I}(e_{0},e_{1})
-\frac{\epsilon}{2} )
\]
belongs to $J^{+}(m_{0k})$. But $m_{1k} \to m_1=(e_1, y_{1})=(e_1,
y_0+ \mathcal{I}(e_{0},e_{1})-\frac{\epsilon}{2} )$ and by causal
simplicity $m_{1} \in J^{+}(m_0)$ which due to Eq. (\ref{kjh}) it is
impossible.

An analogous argument works for $t_0=t_1$. Here we have to consider
two possibilities, $q_0\ne q_1$ and $q_0=q_1$.

In the former case if $\mathcal{I}$ were not lower semi-continuous
at $(e_{0},e_{1})\in E \times E$,  then there would be an $M>0$ and
a sequence $(e_{0k},e_{1k}) \to (e_0,e_1)$ such that $
\mathcal{I}(e_{0k},e_{1k})< M$. Without loss of generality we can
assume $q_{0k} \ne q_{1k}$ and hence necessarily $t_{0k}<t_{1k}$
(otherwise $\mathcal{I}(e_{0k},e_{1k})=+\infty$). Fixed $y_0 \in
\mathbb{R}$ and arguing as above it follows that,
$m_{1k}=(e_{1k},y_0+M)\in J^{+}(m_{0k})$, where
$m_{0k}=(e_{0k},y_0)$. By causal simplicity $m_{1}=(e_1,y_0+M) \in
J^{+}(m_0)$, where $m_0=(e_0,y_0)$ which is impossible because no
causal curve can connect two events in the same slice $t=cnst.$
unless both events lie in the same geodesic generated by $n$ in
which case $q_0=q_1$, which is excluded by assumption.

In the latter case $e_1=e_0$ and if $\mathcal{I}$ were not lower
semi-continuous at $(e_{0},e_{1})\in E \times E$,  then there would
be and $\epsilon>0$ and a sequence $(e_{0k},e_{1k}) \to (e_0,e_1)$
such that $ \mathcal{I}(e_{0k},e_{1k})<
\mathcal{I}(e_{0},e_{1})-\epsilon=-\epsilon$. The inequality implies
that necessarily $t_{0k}<t_{1k}$, and proceeding as above we obtain
that $m_{1}=(e_1, y_0-\frac{\epsilon}{2})=(e_0,
y_0-\frac{\epsilon}{2}) \in J^{+}(m_0)$, $m_0=(e_0,y_0)$, which is
impossible because  the only causal curve which connects $m_0$ to
$m_1$ is a {\em past} directed null geodesic generated by $n$. The
lower semi-continuity for $t_1<t_0$ is obvious because this is an
open subset of $E\times E$ where $\mathcal{I}$ has constant value
$+\infty$.

Now,  assume that properties (i) and (ii) hold and let us prove that
$(M,g)$ is causally simple. Since the Eisenhart spacetime is causal
(lemma \ref{xcv}) we have only to prove that for every $m \in M$,
$J^{+}(m)$ is closed \cite[Def. 3.63]{minguzzi06c} \cite{bernal06b}
(the analogous past case property can be proved similarly).
Otherwise there are two events $m_0=(e_0,y_0)$ and $m_1=(e_1,y_1)$
such that $m_1 \in \bar{J}^{+}(m_0)$ but $m_{1} \notin J^{+}(m_0)$.

If $t_0<t_1$, by the same argument used above, for any  $q \in
C^{1}_{e_0,e_1}$ and $C\ge 0$
\[
(e_{1},y_0+I_{e_{0},e_{1}}[q]+C)
\]
belongs to $J^{+}(m_{0})$. Thus, if
$\mathcal{I}(e_{0},e_{1})=-\infty$, $m_1 \in J^{+}(m_0)$ the
searched contradiction. Let $\mathcal{I}(e_{0},e_{1})$ be finite,
then by property (ii) there is a minimizing curve $(t,\bar{q}(t))$
whose light lift starting at $m_0$ ends at
$\bar{m}_1=(e_1,\bar{y}_1)=(e_1,,y_0+\mathcal{I}(e_0,e_1))$.
Moreover, if $\bar{y}_1 \le y_1$ then $m_1 \in J^{+}(m_0)$  (simply
move from $m_0$ to $\bar{m}_1$ through the light lift of
$(t,\bar{q}(t))$ and then along the fiber passing through $m_1$ in
the forward direction) a contradiction. We conclude that  $y_1 <
\bar{y}_1$. We are going to show that this fact is incompatible with
property (i). Indeed, let $m_{1k}=(e_{1k}, y_{1k}) \in J^{+}(m_0)$
be a sequence such that $m_{1k} \to m_1$. There is an $\epsilon>0$
such that for sufficiently large $k$, $y_{1k}<\bar{y_1}-\epsilon$
then $\mathcal{I}(e_{0},e_{1k })<\mathcal{I}(e_{0},e_{1})-\epsilon$
which contradicts the lower semi-continuity of $\mathcal{I}$.

If $t_0=t_1$ and $q_0\ne q_1$, it is $\mathcal{I}(e_0,e_1)=+\infty$.
Let $m_{1k}=(e_{1k}, y_{1k}) \in J^{+}(m_0)$ be a sequence such that
$m_{1k}=(e_{1k},y_{1k}) \to m_1=(e_1,y_1)$. We can assume without
loss of generality $q_{1k}\ne q_{0}$, and hence $t_{1k}>t_0$
(because all the events of $J^{+}(m_0)$ have time greater than $t_0$
unless they lie in the geodesic generated by $n$ which pass through
$m_0$ in which case they would project on $(q_0,t_0)$). Since
$y_{1k}\to y_1$, there is a $M>0$ such that $y_{1k}<M$ thus
$\mathcal{I}(e_0,e_{1k})<M$, in contradiction with the lower
semi-continuity of $\mathcal{I}$.

If $t_0=t_1$ and $q_0= q_1$, it is $\mathcal{I}(e_0,e_1)=0$. Let
$m_{1k}=(e_{1k}, y_{1k}) \in J^{+}(m_0)$ (in particular $t_{1k}>
t_0$ unless $e_{1k}=e_0$ and $y_{1k}\ge y_0$, in which case $t_{1k}=
t_0$) be a sequence such that $m_{1k}=(e_{1k},y_{1k}) \to
m_1=(e_1,y_1)$. Since $m_{1} \notin J^{+}(m_0)$, it must be
$y_1<y_0-\epsilon$, for a suitable $\epsilon >0$, indeed any event
$m_1$ with $y_1\ge y_0$ is reached from $m_0$ by the future directed
lightlike geodesic generated by $n$ starting from $m_0$. Since
$y_{1k} \to y_1 <y_0-\epsilon$ it is $t_{1k}> t_0$, thus
$y_0+\mathcal{I}(e_0,e_{1k})\le y_{1k}<y_0-\epsilon$ in
contradiction with the lower semi-continuity of $\mathcal{I}$.

%
%

\end{proof}

A nice feature of the theorem is that it relates the causal
simplicity of $(M,g)$ with properties of the classical action
functional alone. There is no direct requirement on the time
dependent metric $a_t$, 1-form field $b_t$ and scalar field
$\mathcal{U}(t,q)$. Roughly speaking (i.e. neglecting condition
(i)), it shows that the property of existence of  minimizers for the
classical action is equivalent to the property of causal simplicity
for Eisenhart's spacetime.

Thanks to the theorem it is quite easy to construct spacetimes which
are not causally simple. As a matter of fact similar examples
already appeared in the literature. In fact, after the discovery by
Penrose \cite{penrose65} that pp-waves  are not globally hyperbolic
Ehrlich and Emch \cite{ehrlich92} showed that they are not even
causally simple (for them $a_{a b}=\delta_{a b}$, $a,b=1,2$, $b_t=0$
and $\mathcal{U}$ is a suitable time dependent quadratic form in
$q^1$ and $q^2$).

In their study of the  general plane waves \cite{flores03,flores06}
Flores and S\'anchez prove (using only differential geometric tools)
that if $a_t$ is independent of time, $(S,a)$ is a complete
Riemannian manifold, $b_t=0$, and $\mathcal{U}$ has a subquadratic
behavior at spatial infinity then,
 (a) the spacetime $(M,g)$ is globally hyperbolic \cite[Theor.
4.1]{flores03} (and hence causally simple), and (b) the action
functional is bounded from below \cite[Lemma 3.3]{flores03}. From
theorem \ref{cas}, point (ii), it follows that under the same
conditions the classical action functional admits a minimizer. We
conclude that only by using tools from global Lorentzian geometry it
is possible to prove results on the existence of minimizers that
were previously obtained by using variational tools such as
Ljusternik-Schnirelman and Morse theories \cite{candela03b}.

Theorem \ref{cas} has an analog in the spacelike dimensional
reduction case \cite{minguzzi06f}. Indeed, it can be shown that a
Lorentzian direct product $M\times H$, with $(M,g)$ Lorentzian and
$(H,h)$ Riemannian manifold, is causally simple if and only if the
base spacetime $(M,g)$ is causally simple, the Lorentzian distance
on the base is upper semi-continuous (the analog of condition (i))
and any two causally related events on $(M,g)$ at finite Lorentzian
distance are connected by a maximizing geodesic  (the analog of
condition (ii)).

\section{Conclusions} \label{conc}

In the first part of the work it has been explained how Eisenhart's
spacetime can be recovered from coordinate independent assumptions,
and how this assumptions relate to the Bargmann structures. In the
process a one-to-one correspondence between Newtonian frames and
Abelian connections on suitable lightlike principal bundles, since
now passed unnoticed, has been proved. It is expected to
considerably simplify the presentation and study of Newton-Cartan
theory.

In the second part a causal version of Eisenhart's theorem was
proved and its connection with a Fermat type principle was
clarified. The concept of  {\em light lift} was introduced and
exploited showing that the last coordinate of the light lift is in
fact the classical action up to a constant. The tools introduced led
us to recognize that  the causal simplicity of Eisenhart's spacetime
is almost equivalent to the property of the existence of minimizers
for the associated classical action.

The bridge between relativistic and non-relativistic physics
provided by the mathematics of lightlike dimensional reduction makes
it possible to use the powerful tools from global Lorentzian
geometry to attack, from a different perspective, some unsolved
problems in classical mechanics. More on this subject will be done
in subsequent works.

\section*{Acknowledgements}
I thank M. Modugno for pointing out some references on Newton-Cartan
theory and M. S\'anchez for a careful reading of the manuscript. I
also thank the referees for their useful criticisms.

\section*{Appendix: The relation with Newton-Cartan theory} \label{die3} There
is no (Newtonian frame independent) way of extending the metric
(\ref{aaa}) to a metric in the whole tangent space to $E$.
Nevertheless, if $a^{a b}$ denotes the contravariant metric
$a^{ab}a_{bc}=\delta^{a}_{c}$ then $a^{a b}$ can be naturally
extended to a degenerate contravariant metric on the whole cotangent
space of $E$, $a^{A B}$, $A,B=0,1,\ldots, d$, by setting $a^{AB}
t_{, B}=0$. This contravariant metric $a^{-1}$ can also be obtained
by pulling back the 1-forms from $E$ to $M$ and by applying to them
$g^{-1}$, i.e. $a^{-1}=\pi_{*}g^{-1}$. A structure of this kind, in
which a rank-d contravariant metric  is annihilated by a 1-form
field $\eta=\psi\dd t$ on a $d+1$ dimensional manifold is called a
{\em flat Galilei structure} \cite{kunzle72}. The flat Galilei
structure is of {\em first order} if  $\dd \eta=0$ on $E$, and in
the spacetime $M$ it coincides with the Bargmann structure condition
$\dd n=0$.

If one tries to include an affine connection $\nabla^{(E)}$ on $E$
that preserves the contravariant metric and $\eta$, (i.e. the
Galilei structure), $\nabla^{(E)}a^{-1}=0$, $\nabla^{(E)}\eta=0$,
then a number of nice theorems arise \cite{kunzle72}. For instance,
the affine connection is torsionless iff $\dd \eta=0$. However, this
condition does not determine the connection completely. Indeed, the
torsionless connections turn out to be in one-to-one correspondence
with the 2-forms on $E$. This set is restricted to the closed
2-forms if a particular {\em Newtonian} condition, that is a
symmetry condition on the Riemann  tensor, is imposed on the
physical affine connections. Finally, the dynamical field equations
$R^{(E)}_{AB}=4\pi \rho \eta_A \eta_B$ for a given scalar matter
density $\rho$ imply that the space section $S_t$ are Euclidean
$a_{a b}=\delta_{a b}$, that the connection only depends on a
potential $U$ in the usual way \cite[Chapter 12]{misner73},
$\Gamma^{(E)\,a}_{\ \ \ 00}=\p_{a}U$,  and that $U$ satisfies the
Poisson equation $ \triangle U=4\pi \rho$. We see that two seemingly
unrelated properties of the old Newtonian spacetime of classical
mechanics, such as Newton's force law and the flatness of space are
unified in the same geometrical framework.

Let us exploit the relation between the Newton-Cartan theory
sketched above and the Bargmann structures \cite{duval85,duval91}.
The metric $g$ on $M$ determines an affine connection $\nabla$ which
projects into an affine connection on $E$ as follows. Let $w$, be a
vector field on $TE$, and let $W$ be any representative on $M$,
$\pi_{*}W=w$. This last condition can be written $L_nW\propto n$,
 and implies from $L_{X}Y=[X,Y]=\nabla_{X}Y-\nabla_{Y}X$, and from the fact
that $n$ is covariantly constant, $\nabla_n W \propto n$. As a
consequence
\[
\nabla_{V+\alpha n} (W+\beta n)= \nabla_{V}W+\alpha \nabla_{n}W+n
(\nabla_{V+\alpha n}\beta)
\]
has a projection independent of the fields $\alpha$ and $\beta$, and
hence defines a connection $\nabla^{(E)}_{v}w=\pi_{*}(\nabla_{V}W)$
on $E$. It is easy to check that $\nabla^{(E)}$ preserves the
Galilei structure. Moreover, it can  be shown to be  Newtonian
\cite{duval85}. The Bargmann structure is then simpler than the
Galilei structure since no Newtonian condition on the Galilei
connection must be added. It appears to be extremely natural for the
formulation of classical theories.

Newton-Cartan theory could prove important at the quantum level
although very few works deal with the problem of formulating it. As
it was pointed out by Kuch$\check{ \rm a}$r \cite{kuchar80} the
quantized Newton-Cartan theory would represent a  face of the {\em
cube of theories} (parametrized with $1/c$, $G$, $\hbar$, and where
classical mechanics corresponds to $(000)$), that is the face of
equation $1/c=0$, the other two known faces being Quantum Field
Theory, i.e. $G=0$, and General Relativity, i.e. $\hbar=0$. The
quantum theory of a particle on a curved Newton-Cartan background
can be approached by means of path integral quantization methods or
through  geometric quantization strategies \cite{modugno00}. Little
is known on the quantization of the geometric fields, but it seems
that they would be non-dynamical \cite{kuchar80}. \\

\end{document}